\documentstyle[12pt,aasms4]{article}

\begin{document}
\title{A comparative HST imaging study of the host galaxies of
radio-quiet quasars, radio-loud quasars and radio galaxies: Paper I}
\author{R.J. McLure}
\affil{Institute for Astronomy, University of Edinburgh, Blackford Hill,
Edinburgh, EH9~3HJ, UK}
\author{J.S. Dunlop}
\affil{Institute for Astronomy, University of Edinburgh, Blackford Hill,
Edinburgh, EH9~3HJ, UK}
\author{M.J. Kukula}
\affil{Institute for Astronomy, University of Edinburgh, Blackford Hill,
Edinburgh, EH9~3HJ, UK 
\& Space Telescope Science Institute, 3700 San Martin Drive, Baltimore,
MD 21218, U.S.A.}
\author{S.A. Baum}
\affil{Space Telescope Science Institute, 3700 San Martin Drive, Baltimore,
MD 21218, U.S.A.}
\author{C.P. O'Dea}
\affil{Space Telescope Science Institute, 3700 San Martin Drive, Baltimore,
MD 21218, U.S.A.}
\and
\author{D.H. Hughes}
\affil{Institute for Astronomy, University of Edinburgh, Blackford Hill,
Edinburgh, EH9~3HJ, UK}

\begin{abstract}
We present the first results from a major HST {\it WFPC2} imaging study 
aimed at providing the first statistically meaningful comparison 
of the morphologies, luminosities, scalelengths and colours of the
host galaxies of radio-quiet quasars, radio-loud quasars, and radio galaxies.
We describe the design of this study and present the images 
which have now been obtained for approximately the first half of
our 33-source sample.
We also report and discuss the results 
of combining high-dynamic range PSF determination with 2-dimensional
modelling to extract the key parameters of the host galaxies of these
AGN; this is the first substantial study of quasar hosts in which it 
is proving possible to determine unambiguously 
the morphological type of the hosts of all 
quasars in the sample.
A full statistical analysis is deferred until the complete sample has been
observed, but within the sub-sample presented here we have already
obtained some remarkably clean results. We find 
that the underlying hosts of all three 
classes of luminous AGN are
massive elliptical galaxies, with scalelengths $\simeq 10$ kpc, and
$R-K$ colours consistent with old stellar populations.
Most importantly this is the the first unambiguous evidence that,
just like radio-loud quasars, 
essentially all radio-quiet quasars brighter than $M_R = -24$ 
reside in massive ellipticals.
This result removes the possibility that radio `loudness' is directly 
linked to host galaxy morphology, but is 
however in excellent accord with the
black-hole/spheroid mass correlation recently highlighted by Magorrian
{\it et al.} (1998). To demonstrate this we apply the spheroid
luminosity/spheroid mass/black-hole mass 
relations given by Magorrian
{\it et al.} to
infer the expected Eddington luminosity (and hence maximum expected nuclear
$R$-band luminosity) of the putative black hole at the
centre of each of the spheroidal host galaxies we have uncovered.
Comparison of the predicted Eddington 
luminosities with the actual nuclear $R$-band luminosities 
produces a clear relationship,
and suggests that the black holes in most of these galaxies are radiating
at a few percent of the Eddington luminosity (although a few
appear to be radiating close to the Eddington limit); the 
brightest host galaxies in our low-redshift sample are capable of
hosting quasars with $M_R \simeq -28$, comparable to the most luminous
quasars at $z \simeq 3$, if fueled at the Eddington rate.
Finally we discuss our host-galaxy-derived black-hole
masses in the context of the 
radio-luminosity:black-hole mass correlation 
recently uncovered for 
nearby galaxies by Franceschini {\it et al.} (1998). The
radio loud AGN (both RGs and RLQs) lie on the $P_{5GHz}^{total} 
:m_{bh}$ relation, while the RQQs lie on the $P_{5GHz}^{core}
:m_{bh}$ relation. This finding may hold the key to
identifying the physical origin of the radio-loud radio-quiet dichotomy.
\end{abstract}

\section{Introduction}

Studies of the host galaxies of active galactic nuclei (AGN) may hold
the key to answering several fundamental questions about these still poorly
understood objects. Such questions include i) what is the physical origin of radio
loudness?, ii) by what mechanisms are galactic nuclei
triggered into activity ({\it e.g.} Smith \& Heckman 1990; Hutchings \&
Neff 1992; Baum, Heckman \& van Breugel 1992)?, iii)
which classes of AGN can be unified via orientation effects (Peacock 1987;
Barthel 1989; Urry \& Padovani 1995), iv) which classes of AGN can or cannot be unified via
time evolution ({\it e.g.} Ellingson, Yee \& Green 1991; Baum, Zirbel \&
O'Dea 1995), and how does the
recently inferred relation between black-hole mass and galaxy bulge mass (Magorrian {\it et
al.} 1998) extend to high masses?
Furthermore, by defining the parameter space occupied by AGN hosts,
important constraints can be derived on the fraction of the galaxy
population which might contain a dormant AGN, constraints which need to
be satisfied by any physical model which endeavours to explain the cosmological
evolution of active galaxy populations (Dunlop 1997; 
Small \& Blandford 1992; Haehnelt \& Rees 1993, Silk \& Rees 1998)

Determining the properties of the hosts of relatively low-luminosity AGN
such as Seyfert nuclei has proved relatively straightforward since the
advent of CCD detectors, and their
properties are now relatively well established ({\it e.g.} MacKenty 1990). 
However, to date
the effective ground-based 
study of the hosts of the more luminous quasars ({\it e.g.} Smith {\it et
al.} 1986; V\'{e}ron-Cetty \& Woltjer 1990) has proved
to be extremely difficult due to the combination of their larger
cosmological distances, and much higher nuclear:host luminosity ratios.
This is unfortunate because a clear understanding of both the differences and
similarities between the host galaxies of the three main classes of
{\it powerful} active galaxy - radio-quiet quasars (RQQs), radio-loud quasars
(RLQs) and radio galaxies (RGs) - is obviously of fundamental importance in any
attempt to unify or relate the various manifestations of the AGN
phenomenon.

One way to reduce the distorting influence of the nuclear light is to
observe at near-infrared wavelengths where the nuclear:host luminosity
ratio can be minimized. We have previously exploited this fact, while at
the same time attempting to minimize cosmological distance, via an
extensive $K$-band imaging study of the hosts of matched samples of
RQQs, RLQs, and RGs in the redshift range $0.1 \leq z < 0.35$ 
(Dunlop {\it et al.} 1993; Taylor {\it et al.} 1996).
This infrared study was successful in that, unlike previous
ground-based optical studies, it proved possible to determine reliably
the morphological type of a substantial fraction ($> 0.5$) of the quasar
hosts. Moreover, the use of properly matched samples allowed us to
demonstrate that the hosts of all three classes of powerful AGN were
large ($\ge 10$ kpc), luminous ($> 2 L^{\star}$) galaxies, and that the
hosts of RLQs and RGs were statistically identical, consistent with
unified models. We also found that at least some of the RQQs appeared to
lie in elliptical rather than disc-like galaxies, and there was a
suggestion in our data that the probability of an RQQ having an
elliptical host is an increasing function of quasar luminosity. However
we were unable to prove this effect was significant due to the fact that
the morphologies of the hosts of the more luminous RQQs in our sample
remained ambiguous with the limitations of ground-based seeing.
This drawback, coupled with the desirability of obtaining reliable optical-infrared colours 
for all the host galaxies in our AGN sample led us to undertake a
complementary $R$-band imaging study of this same sample with WFPC2 on
the HST. Here we report the first results of this, the most detailed HST
study to date of the host galaxies of quasars and radio galaxies.

While the observed nuclear:host luminosity ratio of a quasar is inevitably
much higher at $R$ than at $K$, the high spatial resolution offered by
HST allows the nuclear contamination to be confined to the central
regions of the host galaxy image, thus enhancing the prospects of
reliable determination of host galaxy morphology, luminosity and
scalelength. This resolution advantage has previously been explored in a number of
pilot HST studies of small samples of quasars (Disney {\it et al.} 1995; 
Bachall, Kirkhados \& Schneider 1994,1995a,b,c; Hutchings {\it et al.}
1994; Hutchings \& Morris 1995) although it comes with the 
price that the nucleus inevitably
saturates in any HST image of sufficient depth to produce a useful image
of the underlying host. As a result it has proved difficult to perform accurate
subtraction of the nuclear contribution, which is still vitally important
for the reliable determination of host galaxy properties (Hutchings 1995).
A second problem with some previous HST studies has been
the use of filters which have included strong emission lines, making it
difficult to discern which host galaxy features can be reliably
attributed to starlight (Bachall, Kirkhados \& Schneider 1995a). 
A third problem has been the use of the wide
$V$-band filter, which fails to properly sample the dominant stellar population of
the host (at $\lambda_{rest} > 4000$\AA) for $z > 0.25$ (Bachall,
Kirkhados \& Schneider 1994).
A fourth problem
has been the unavailability of an accurate point spread function (PSF) of
sufficient depth to investigate the contribution of the quasar nucleus at
large radii ($> 3$ arcsec) due to the problem of scattered light within WFPC2.

As described in more detail below, our new 34-orbit HST $R$-band study of the
hosts of RQQs, RLQs and RGs has been designed to overcome these problems
and differs from previous HST studies of quasar
hosts in five important ways. 
First, we are imaging statistically comparable
samples of each class of AGN. Second, we already possess deep infrared
images of all our targets, which will allow the first meaningful study of
the optical-infrared colours of quasar hosts. Third, we have used the F675W
$R$-band filter, and restricted the redshift range of our targets, in
order to ensure that the images are always uncontaminated by emission
lines, and always sample the rest frame emission of the host galaxy
longward of the 4000\AA\ break. Fourth, we have devoted an orbit of our HST 
observing programme to assembling an accurate point spread function of
sufficient dynamic range to accurately 
define the contribution of the quasar nucleus out to an angular 
radius $r > 10$ arcsec.
Fifth, we have developed and applied a 2-dimensional modelling procedure 
which allows us to reliably extract the morphology, luminosity and size 
of the host galaxies from our images, without requiring us to make
asumptions {\it a priori} about the values of these parameters (which
contrasts with the analysis of Bachall, Kirkhados \& Schneider (1994) as
highlighted by McLeod \& Rieke (1995)).

The results presented here 
from the first year of this study demonstrate the importance of
these five  improvements. In particular, whereas most previous HST studies have
tended to highlight the fact that many quasar hosts display a wide range of morphological
peculiarities, our study is already revealing a surprising degree of
similarity and homogeneity in the distribution and age of the dominant
stellar populations in the hosts of these powerful AGN.

The layout of the paper is as follows. In section 2 we summarize the main
properties of the matched RG, RLQ and RQQ samples which are the subject
of this HST study, and then in section 3 we give details of the
observations, image reduction and PSF determination. In section 4 we
present the images, list and summarize the results of applying our
host-galaxy modelling procedure to these data, and provide brief notes
on the images of each of the 19 AGN observed to date in the context of
previous observations. In section
5 we discuss the main implications of the trends uncovered by the
initial results of this study, and in section 6 we summarize our
principle conclusions. Unless otherwise stated $\Omega_0 = 1$ and $H_0 =
50$ ${\rm km s^{-1} Mpc^{-1}}$ are assumed throughout, and we convert 
previously published scalelengths and luminosities to this cosmology for
ease of comparison.
   
\section{The sample}

The full sample selected for HST imaging consists of 33 objects (10 RLQs, 12 RQQs and 11 RGs) selected
from the slightly larger statistically-matched samples which were imaged in
the near-infrared by Dunlop {\it et al.} (1993) and Taylor {\it et al.} 
(1996). Full
details of these samples can be found in these papers. The key point
is that the RLQ and RQQ sub-samples were selected to be statistically 
indistinguishable in terms of optical luminosity and redshift, while the
RLQ and RG sub-samples were selected to be indistinguishable in terms of
radio luminosity, radio spectral index and redshift. 
Note that following the radio observations of Kukula {\it et al.} (1998a)
we now possess either a radio detection or a strong upper limit on radio
luminosity for all the RQQs in our sample, making it possible to quantify
their `radio quietness', and hence explore the extent to which this can
be related to any properties of the host galaxy (see Section 5).

The original combined
sample comprised a total of 40 objects (12 RGs, 13 RLQs and 15 RQQs) with
$0.1 < z < 0.35$, but for this HST study we have restricted the
redshift range in order to avoid 
[O{\sc iii}] emission entering the blue end of the F675W filter. This results in the slightly smaller sub-samples described
above, without compromising their statistical compatibility. 
The 19 objects which have been observed during the first year of this
study, and for which we present the data in this first paper, are listed
in table 1, along with dates on which they were observed with the HST.
 
\section{Observations}

\subsection{Detector and filter choice}
Observations were made with the Wide Field \& Planetary Camera 2 ({\it
WFPC2}; Trauger {\it et al.} 1994) on the Hubble Space Telescope using the F675W filter. The
filter spans 877\AA~ in the wavelength range 6275.5 -- 7152.5\AA,
roughly equivalent to standard $R$ band, and thus excludes both [O{\sc
iii}] $\lambda 5007$ and H$\alpha$ emission for redshifts $0.1\leq z
\leq 0.25$ (a wider filter, although providing greater throughput,
would have allowed our images to become contaminated by line emission
which could mask or at least be confused with 
the underlying stellar continuum of the host galaxy).

{\it WFPC2} consists of four detectors, each comprising 
$800\times800$ pixels:
three WF chips, each with a pixel scale of 100~mas; and one PC chip,
with a pixel scale of 45~mas.  Although the PC chip offers smaller
pixels and correspondingly better sampling of the instrument point
spread function (PSF), it is ultimately less sensitive to low surface
brightness emission than the WF chips, even when the pixels are binned
up, and we therefore opted to use the larger detectors.  Target
sources were centred on the WF2 chip, which was chosen for its
marginally better performance over the period immediately prior to our
observations.

\subsection{Observing strategy and image reduction}

Observations of the target quasars and radio galaxies were carefully
tailored to ensure that the maximum amount of information could be
derived from the final images.  In both cases deep, sensitive images
of the galaxies are clearly desirable, but for the quasars such
exposure times inevitably entail saturation of the central source,
allowing no independent measure of PSF normalization.

Slightly different strategies were therefore used for the quasar and
radio galaxy samples. For the quasars, exposures of 5, 26 and
$3\times600$ seconds were taken. The short exposures guaranteed that
at least one unsaturated image of the quasar would be obtained, thus
ensuring an accurate measure of the  
central flux density. The three 600-second
exposures each provided a $3\sigma$ surface brightness 
sensitivity $\mu_R = 23.8$ mag.
arcsec$^{-2}$ (per pixel), and their comparison
facilitated reliable cosmic ray removal using standard {\sc iraf} tasks.
With azimuthal averaging, the combined 1800-second deep image of each
quasar allows extended emission to be traced reliably out to a surface
brightness level $\mu_R > 26$ mag. arcsec$^{-2}$.

For the radio galaxies there was little danger of saturation and so short
exposures were not required.
Three 700-second exposures were therefore obtained for each radio galaxy.  
Any remaining time in the orbit was filled with an exposure of flexible 
length (usually 40 to 100 seconds).

The sources discussed in this paper - constituting approximately half
of our sample - were all observed between June 1997 and April 1998
(see table 1 for exact dates). Calibration was carried out using the
standard pipeline.

\subsection{Determining the point spread function}

The form of the {\it WFPC2} point spread function depends critically
on both the position on the chip and the spectral energy distribution
(SED) of the target source. These effects are well-understood and can
be included in software to produce accurate synthetic PSFs.  However,
despite providing an excellent fit over the central few arcseconds, the
synthetic PSFs produced by packages such as {\sc tinytim} deviate from
the empirical {\it WFPC2}\, PSF at larger radii ($\ge 2$~arsec). This
is due to scattering within the camera. Since the scattered light
shows complex structure which is not uniform, and which is also
wavelength and position dependent, it cannot easily be modelled.

We have therefore devoted one orbit of our allotted HST time to constructing
a deep, unsaturated stellar PSF using the F675W filter, with the star
centred on exactly the same part of the WF2 chip as the target objects.

The star chosen was GRW~+70D5824, a white dwarf of spectral type DA3
and apparent magnitude $V=12.77$. Since it also serves as a UV
standard star for {\it WFPC2} the position and spectrum of this object
are extremely well determined (Turnshek {\it et al.} 1990). No stars of a
comparable brightness lie within 30~arcsec, ensuring that the stellar
PSF is not contaminated by PSFs or scattered light from neighbouring
objects.  The $B-V$ colour of the PSF star is -0.09, sufficiently
similar to the neutral colours ($B-V \simeq 0$) typical of our quasar
sample to provide a reasonable match to a quasar SED over the
wavelength range of the F675W filter.

In order to obtain both unsaturated images of the PSF core {\it and}
deep images of the wings, a series of exposures was carried out with
durations of 0.23, 2, 26 and 160 seconds. After 0.23 seconds the central
pixel of the stellar image reaches approximately 10\% of its
saturation value (assuming that the star is perfectly centred). This
is the shortest practical exposure length - in exposures of less than
0.23 seconds the PSF would be compromised by the shutter flight time.
Subsequent exposure durations were carefully staggered to ensure that
the star never saturated beyond the radius at which the wings of the
PSF in the previous, shorter exposure became lost in the noise. This
allowed us to build up a composite PSF of very high dynamic range by
splicing together annuli from successively deeper exposures.

The brightest quasar in our sample has an apparent magnitude of
$V=15.15$, more than two magnitudes fainter than the PSF star. We were
therefore able to ensure that the deepest stellar image probed much
further into the wings of the PSF than even the longest quasar
exposure.

Each exposure also used a two-point dither pattern to improve the
sampling of the PSF (which is significantly undersampled by the
0.1~arcsec pixel scale of the WF chips).

Thus, the stellar PSF is designed to match the PSFs of the target
quasars as closely as possible in terms of depth, position on the
detector and spectral energy distribution. However, we note that we cannot account for time-dependent variations in the PSF using this method. These variations are due to changes in the
telescope focus and include contributions from several sources. Of these
the most significant for our observations is the short-term (intra-orbit) 
variability (`breathing') due to temperature fluctuations in the
telescope's environment. However, with the relatively large pixel scale of
the WF chips the effect on the amount of flux falling on the central pixel
of a point source is likely to be only a few percent at most.

\section{RESULTS}

\subsection{Images}

The images, two-dimensional model fits, and model-subtracted residual
images are presented in figures 1a -- 1s.
A grey-scale/contour image of the final reduced 
F675W $R$-band image of each AGN 
is shown in the top
left panel (panel A) of each figure 1a$-$1s, 
which shows a region 12.5 $\times$ 12.5 arcsec
centred on the target source. The surface brightness of the lowest contour 
level is indicated in the top-right corner of the panel with the greyscale 
designed to highlight structure close to this limit.  Higher surface brightness 
contours are spaced at intervals of 0.5 mag. arcsec$^{-2}$, and have been
superimposed to emphasize brighter structure in the centre
of the galaxy/quasar.  Panel B in each figure shows the best-fitting 
two-dimensional model, complete with unresolved nuclear component (after
convolution with the empirical PSF) contoured in an identical manner 
to panel A. 
Panel C shows the best-fitting host galaxy as it would appear if 
the nuclear component were absent, while panel D is the residual image 
which results from subtraction of the full two-dimensional model (in
panel B) from the raw $R$-band image (in panel A), in order to highlight
the presence of morphological peculiarities such as tidal tails,
interacting companion galaxies, or secondary nuclei.
All panels are displayed using the same greyscale.

\subsection{Modelling Results}

Full details of the 2-dimensional modelling procedure which we have
used to determine the properties of the host galaxies are presented
elsewhere, along with the results of extensive tests of its ability to
reclaim the true properties of a wide range of host-galaxy:nucleus
combinations at different redshifts (McLure et al. 1998b). In brief, the
modelling procedure is a development of that used by Taylor et al.
(1996) and here we have used two distinct versions of this procedure to
determine the host-galaxy properties from our HST images. For both versions
an accurate high-dynamic range PSF and an accurate error frame for each
quasar image are essential for the extracton of robust results (see
McLure et al. 1998b for details).

In the first
version the host galaxy morphology is constrained by determining how
well the data can be reproduced {\it assuming} that the host galaxy is 
{\it either} an elliptical galaxy (described by a de Vaucouleurs $r^{1/4}$-law) 
or an exponential disc. The remaining five parameters (host-galaxy position
angle, host-galaxy axial ratio, host-galaxy scalelength, host-galaxy
luminosity and nuclear luminosity) are then varied until, when convolved
with the PSF, the model best fits the data as determined by
$\chi$-squared minimization (note that it is not assumed {\it a priori} 
that the radio galaxies have a negligible nuclear component). 
Then, if one assumed galaxy morphology
yields a significantly better fit than the other, we can say that the
galaxy is {\it better} described by a de Vaucouleurs law or by an
exponential disc. The results of applying this procedure to the HST
images are given in table 2. The striking feature of
these results is that all of the host galaxies except those of the two lowest
luminosity RQQs are better described as elliptical galaxies, and that,
with only one exception (the heavily nuclear dominated RQQ 0953$+$415), this
difference is statistically very significant.
One dimensional luminosity profiles extracted from the 
best-fitting de Vaucouleurs or exponential models are compared
with the data in figure 2.

In the second version we have removed the need to assume that the host galaxy
can be decribed by either a pure $r^{1/4}$-law or exponential disc, and
allow a sixth parameter $\beta$ (where the luminosity profile of 
profile of the galaxy is given by $I(r) \propto exp(-r^{\beta})$ )
to vary continuously. Thus, $\beta = 1$ should result if the galaxy
is best described by a pure exponential disc, and $\beta = 0.25$ should
result if the galaxy really does follow a pure de Vaucouleurs law,
but {\it all} 
values of $\beta$ are available to the program to improve the quality
of the model fit. The impressive results of applying this procedure to the HST
images are given in table 3 and illustrated in figure 3. One remarkable 
feature of these results (see figure 3) is that for 16 out of the 
19 objects, the preferred $\beta$
parameter for the host galaxy is in the range $0.18 < \beta < 0.3$, in
excellent agreement with a pure de Vaucouleurs law. Perhaps most
surprisingly, given pre-existing prejudices, this is true for 6 out of the 9
RQQs in the current sub-sample.  Moreover, the only two
RQQs which appear to have a significant disc component are the two least
luminous RQQs in this subsample. The clear implication is that {\it all}
 bright quasars, with $M_R < -23.5$ reside in massive ellipticals,
irrespective of their radio power. Such a result has been hinted at
before (Taylor {\it et al.} 1996; Disney {\it et al.} 1995) 
but this is the first time 
it has proved possible to demonstrate unambiguously that this is the case.
The modelling results are discussed further in Section 5.

\subsection{Notes on Individual Objects}

Here we provide a brief discussion of the HST image of each 
object presented in this paper, with reference to other recent HST 
and ground-based data.  A more detailed description of each object, 
together with our existing $K$--band images, can be found in 
Dunlop {\it et al.} (1993) and Taylor {\it et al.} (1996).  Sources 
are listed by IAU name, with alternative names
given in parenthesis.  Radio luminosities or upper limits at 5 GHz 
have been calculated assuming $H_0 = 50 {\rm km s^{-1} Mpc^{-1}}$ and
$\Omega_0 = 1$. Note that following the deep VLA radio observations 
of RQQs undertaken by Kukula {\it et al.} (1998a), radio detections or
strong upper limits are now available for all the RQQs in the current
sample.

\subsubsection{The Radio Galaxies}

\vspace*{0.2in}

\noindent
{\bf0345$+$337} (3C 93.1, 4C $+$33.08, B2 0345$+$33, NRAO 0146, DA 113, OE $+$376)

\vspace*{0.07in}

\noindent
($z=0.244$, Log$_{10}(L_{5GHz}/{\rm W Hz^{-1} sr^{-1}})=25.45$)

\vspace*{0.07in}

\noindent
The results of our modelling of this object, shown in figure 1a, reveal 
the host to be a 
large elliptical galaxy, with $r_{1/2}=11$ kpc.  The strong preference 
for a elliptical host is confirmed by the variable-beta modelling, which
yields $\beta=0.249$. 
The unresolved central point source is weak, 
making up $\approx6\%$ of the integrated flux in the best fitting model.  
The model-subtracted image shows some low-level residual flux  unaccounted 
for by the symmetrical host template. 

\noindent
The embedded companion detected 6\,arcsec NW of the main galaxy in our $K$--band image is seen as clearly separated in the new $R$--band image.  The other companion detected to the SE at $K$ is again detected here, together with several other faint companion objects.  The two steep isophote companions are confirmed to be foreground stars from our HST image.   

\noindent
At high resolution in the radio this is a compact steep-spectrum source
having a diameter of $\simeq0.4$\,arcsec (Akujor {\it et al.} 1991).

\vspace*{0.2in}

\noindent
{\bf0917$+$459} (3C 219, 4C $+$45.19, NRAO 0320, DA 266, LHE 249, OK $+$430)

\vspace*{0.07in}

\noindent
($z=0.174$, Log$_{10}(L_{5GHz}/{\rm W Hz^{-1} sr^{-1}})=25.69$)

\vspace*{0.07in}

\noindent
A disc host for this radio galaxy is excluded by our model fitting with a 
high level of confidence.  As shown in figure 1b, 
the host is a large elliptical, $r_{1/2}=11$ kpc, 
with a very weak, unresolved nuclear component.  
The variable-beta model again chooses an almost perfect 
de Vaucouleurs template, with $\beta=0.229$.  This galaxy has been 
recently imaged with HST in a snapshot survey of 3CR radio galaxies 
(De Koff {\it et al.} 1996), through the F702W (wide $R$) filter.  
De Koff {\it et al}\, suggest that there may be an interaction with the 
large galaxy to the SE.  However, our model-subtracted image provides
little direct evidence for any interaction.

\noindent
This galaxy is clearly situated in a cluster, with a large number of 
companion objects detected at $R$ and in our previous $K$--band image.  
In the radio this is a classical double source whose position angle is 
anti-correlated with the optical and near-infrared position angles. 

\vspace*{0.2in}

\noindent
{\bf0958$+$291} (3C 234.0, 4C $+$29.35, IRAS F09589$+$2901, B2 0958$+$29,
NRAO 0343,
CSO 0031, OL $+$200, DA 280, CTD 064, CTA 049)

\vspace*{0.07in}

\noindent
($z=0.185$, Log$_{10}(L_{5GHz}/{\rm W Hz^{-1} sr^{-1}})=25.30$)

\vspace*{0.07in}

\noindent
As shown in figure 1c, the host is extremely well fitted by an 
elliptical galaxy template, 
with $r_{1/2}=8.3$ kpc, and a disc host is formally excluded.  
This result is supported by the variable-beta model, which yields another
virtually perfect de Vaucouleurs model ($\beta=0.253$). 
This galaxy has a rather more luminous 
unresolved component, which contributes nearly $30\%$ of the integrated flux.
3C234.0 was also included in the HST snapshot survey 
(De Koff {\it et al} 1996), where features were detected emanating to the 
east and west of the galaxy nucleus.  The large tidal arm to the west of the 
nucleus is easily visible in the contour plot of figure 1c.\, and is dramatically 
highlighted in the model subtracted image.  A fainter counter arm to the 
east of the nucleus is also present, although it is not easily discerned
in this greyscale image.

\noindent
Numerous companion objects are detected in our $R$--band image in 
agreement with the ground-based images of Hutchings, Johnson \& Pyke (1988) 
through the same filter.  Recent work by Young {\it et al} (1998) has led
to the detection of broad $H\alpha$ in both total and polarized flux, 
consistent with our discovery of significant nuclear emission at $R$.
In the radio 3C234 is a classical double FRII source (Leahy, Pooley 
\& Riley 1986).

\vspace*{0.2in}

\noindent
{\bf2141$+$279} (3C 436, 4C $+$27.47, B2 2141$+$27B, NRAO 0665, CTD 132, DA 559, CTA
096)

\vspace*{0.07in}

\noindent
($z=0.215$, Log$_{10}(L_{5GHz}/{\rm W Hz^{-1} sr^{-1}})=25.17$)

\vspace*{0.07in}

\noindent
A large elliptical host galaxy ($r_{1/2}=21$ kpc) is strongly favoured 
for this object, with the variable-beta model again yielding an almost
perfect de Vaucouleurs law ($\beta=0.246$), with an insignificant central
point source contribution.
As can be seen in Panel D of figure 1d, a 
secondary nucleus lies approximately 0.6 arcsec from the
centre of the host galaxy, and this can be clearly seen in the 
model-subtracted image along with some asymmetrical residual flux, and 
several faint companion objects.  
The radio source has an FRII morphology (McCarthy {\it et al.}
1991).

\subsubsection{The Radio Loud Quasars}

\vspace*{0.2in}

\noindent
{\bf0137$+$012} (PKS 0137$+$012, PHL 1093, 4C 01.04, OC 062, UM 355)

\vspace*{0.07in}

\noindent
Radio Loud; ($z$ = 0.258, 
$\log_{10}(L_{5GHz}/{\rm W Hz^{-1} sr^{-1}})$ = 25.26) 

\vspace*{0.07in}

\noindent
The modelling results for this object show the host galaxy to be a 
large elliptical ($r_{1/2}=13$ kpc) which contributes 
over half of the integrated flux at $R$.  The variable-beta model fit is 
close to elliptical ($\beta=0.185$) but does provide a 
significantly better fit than a pure de Vaucouleurs law.  
The contour plot in figure 1e. clearly shows a close companion object 
(separation $\approx 1.0$ arcsec), which was not 
resolved in our $K$-band image.  
This object was masked out during the model fitting and can clearly 
be seen in the subtracted image.  0137$+$012 
was one of four objects imaged with the PC camera on 
HST by Disney {\it et al} (1995) using the F702W filter. 
Using a two-dimensional cross-correlation 
modelling technique they also found the host to be a large early-type 
galaxy.  The slightly larger scale length 
($r_{1/2}\simeq22$ kpc)\footnote{converted to our cosmology} 
and fainter host magnitude obtained by their modelling technique is most
likely to be due
to their use of a synthetic point spread function (PSF).  For 
a more detailed discussion of the problems associated with 
two-dimensional modelling and the WFPC PSF see McLure {\it et al} 
(1998b).  In the radio the quasar is an FRII source of
diameter 42 arcsec with a strong core (Gower \& Hutchings 1984a).

\vspace*{0.2in}

\noindent
{\bf0736$+$017} (PKS 0736$+$01, OI 061)

\vspace*{0.07in}

\noindent
Radio Loud; ($z$ = 0.191, $\log_{10}(L_{5GHz}/{\rm W Hz^{-1} sr^{-1}})$ = 25.35)

\vspace*{0.07in}

\noindent
An elliptical host galaxy is again strongly preferred 
for this quasar, with a disc host formally excluded.  
As with 0137$+$012 the variable-beta 
modelling confirms the generally spheroidal nature of the host 
($\beta=0.1933$) but does provide a statistically significant 
improvement over the pure de Vaucouleurs model. 
The companion object seen  to the south of the quasar, 
first detected in our $K$--band image 
(Dunlop {\it et al} 1993), is not obvious in our new HST image
shown in figure 1f.  
The large area of low surface nebulosity to the north east of the quasar 
which was present in our $K$--band image is also missing from this $R$--band 
image, casting doubt on its reality.  
At radio wavelengths this quasar is a compact (0.013 arcsec) flat-spectrum
source (Romney {\it et al.} 1984; Gower \& Hutchings 1984b).
\vspace*{0.2in}

\noindent
{\bf1004$+$130}(PKS 1004$+$13, PG 1004$+$130, 4C 13.41, OL 107.7)

\vspace*{0.07in}

\noindent
Radio Loud; ($z$ = 0.240, $\log_{10}(L_{5GHz}/{\rm W Hz^{-1} sr^{-1}})$ = 24.94) 

\vspace*{0.07in}

\noindent
As can be seen from the contour plot in figure 1g. this quasar is a 
highly nuclear-dominated object in the $R$-band.  
The preferred host galaxy is again an elliptical 
although due to the high\, $L_{nuc}/L_{host}$\, ratio the 
preference is less clear cut than for most of the other objects.
The best-fitting host is fairly large ($r_{1/2}=8$ kpc) 
and luminous, although the unresolved nuclear component contributes 
$\approx85\%$ of the integrated flux.  The results of the variable-beta 
modelling strongly support the choice of an early-type host 
($\beta=0.253$). 1004$+$130 is one of the quasars which has also been
imaged at the HST in the $V$-band by Bahcall {\it et al.} (1997). 
They also find the host to be best described by an early-type 
galaxy, although with a somewhat smaller scale length ($r_{1/2}=5.8$ kpc).  
They suggest that there is some sort of structure close into the 
quasar nucleus.  This is beautifully illustrated by our model-subtracted 
image; two spiral-arm-type features can be 
clearly seen on either side of the quasar nucleus.  The 
subtracted image also reveals one companion object to the north-east 
together with two fainter companions to the east.  
In the radio this quasar is an FRII source of diameter 9 arcmin with a weak
core (Miley \& Hartsuijker 1978).
\vspace*{0.2in}

\noindent
{\bf2141$+$175} (OX 169, MC3)

\vspace*{0.07in}

\noindent
Radio Loud; ($z$ = 0.213, $\log_{10}(L_{5GHz}/{\rm W Hz^{-1} sr^{-1}})$ = 24.81)

\vspace*{0.07in}

\noindent
Our $R$--band image of this complex object shown in figure 1h 
reveals the two extended 
filaments to the SE and NW previously detected by 
Smith {\it et al} (1986) and Heckman {\it et al} (1986).  The 
best-fitting host galaxy is a moderate sized elliptical 
with $r_{1/2}=4$ kpc, with the variable-beta modelling producing a 
very similar fit. This quasar has been previously imaged with the HST
with the F702W filter by Hutchings {\it et al.} (1994), who also
concluded in favour of an $r^{1/4}$-law but fail to provide any
quantitative information on the scalelength of the host galaxy.
The model-subtracted image shows the 
NW filament to be more extended than previously thought, 
stretching to more than half the length of the SE filament.  
The hypothesis that the two extensions are composed of old stars 
(Stockton \& Farnham 1991) is supported by their prominence in our existing 
$K$-band image (Dunlop {\it et al}\, 1993), and in this emission-line free 
HST image.  
Also revealed by subtraction of the best-fitting model is a previously 
undetected companion $\approx 3$ arcsec to the NE of the quasar.  In 
the radio 2141$+$175 is a compact ($<$ 4 arcsec) 
flat-spectrum source (Fiegelson, Isobe \& Kembhavi 1984).

\vspace*{0.2in}

\noindent
{\bf2247$+$140} (PKS 2247$+$14, 4C 14.82, OY 181)

\vspace*{0.07in}

\noindent
Radio Loud; ($z$ = 0.237, $\log_{10}(L_{5GHz}/{\rm W Hz^{-1} sr^{-1}})$ = 25.31)

\vspace*{0.07in}

\noindent
This radio-loud quasar can be seen to be elongated in the NW/SE direction
in figure 1i, as 
previously reported by Hutchings {\it et al} (1988) and 
Dunlop {\it et al} (1993).  The host galaxy is unambiguously elliptical, 
with a best fitting scale length of $11$ kpc.  
This preference is again impressively 
confirmed by the best-fit beta value of 
$\beta=0.249$.  The subtracted image reveals two 
new companion objects embedded in the residual flux of 
the SE elongation.  The quasar is a compact steep-spectrum 
radio source (van Breugel, Miley 
\& Heckman 1984).
\vspace*{0.2in}

\noindent
{\bf2349$-$014} (PG 2349$-$014, PKS 2349$-$01, PB 5564)

\vspace*{0.07in}

\noindent
Radio Loud; ($z$ = 0.173, $\log_{10}(L_{5GHz}/{\rm W Hz^{-1} sr^{-1}})$ = 24.86)

\vspace*{0.07in}

\noindent
The $R$-band image of this quasar presented in figure 1j, 
shows it to be undergoing an extensive 
interaction.  2349$-$014 is included in the sample of Bahcall 
{\it et al} (1997) (see also Bahcall {\it et al.} 1995a) 
and we confirm their detection of a compact close 
companion at $\approx 2$ arcsec separation to the east.  
Bahcall {\it et al} claim that there is no clear evidence for a normal 
host galaxy centred on the quasar, although they comment 
that the mean radial profile is well matched by a de Vaucouleurs law.  
The results of our two dimensional modelling do not support this.  
Model-fitting after masking of the most prominent areas of asymmetric 
nebulosity produces a good match with 
a large elliptical host galaxy of scalelength $r_{1/2}=18$ kpc.  The 
variable-beta parameter modelling produces a near identical fit 
($\beta=0.258$). The subtracted image highlights the massive tidal arm 
feature to the north of the quasar and extensive nebulosity to the west.  
The source of this interaction would appear to be the nearby 
compact companion.  In agreement  with Bahcall {\it et al} our 
new $R$-band data provides little evidence for interaction with the 
large galaxy to the SE.  In the radio this quasar is an FRII source of diameter
53 arcsec with a strong core (Antonucci 1985).

\subsubsection{The Radio Quiet Quasars}

\vspace*{0.2in}

\noindent
{\bf0054$+$144} (PHL 909)

\vspace*{0.07in}

\noindent
Radio Quiet; ($z$ = 0.171, $\log_{10}(L_{5GHz}/{\rm W Hz^{-1} sr^{-1}})$
= 21.87)

\vspace*{0.07in}

\noindent
The host galaxy of this radio-quiet 
quasar (figure 1k) is extremely well described by an 
elliptical template with $r_{1/2}=8$\ kpc, and a disc host is 
formally excluded.  The variable-beta modelling again strongly supports 
this unambiguous choice, 
settling on a value of $\beta=0.251$. 
This object was also imaged with the HST at $V$ by Bahcall {\it et al.} 
(see Bahcall {\it et al.} (1996)) and in this case they 
also found the best-fitting host to be an early type galaxy.  
In their $V$--band image Bahcall {\it et al.} 
claim not to detect the extended emission towards the western companion 
galaxy (off this frame) which was reported in 
Dunlop {\it et al} (1993).  However, the model-subtracted image presented here 
clearly shows considerable residual luminosity in the direction of the 
western companion in agreement with the $K$-band data.

\vspace*{0.2in}

\noindent
{\bf0157$+$001} (PG 0157$+$001, Mkn 1014)

\vspace*{0.07in}

\noindent
Radio Quiet; ($z$ = 0.163, $\log_{10}(L_{5GHz}/{\rm W Hz^{-1} sr^{-1}})$
= 22.87)

\vspace*{0.07in}

\noindent
This spectacular object can clearly be seen to be undergoing massive tidal 
disruption.  The bright companion object detected at the end of the 
NE tidal arm in our previous $K$-band image is again detected here but 
lies just off the edge of the frame in figure 1l.  Also detected in the 
$R$-band is a faint counter arm to the west, embedded in which are 
two further bright companion objects. The underlying host galaxy is 
best described by a bright elliptical galaxy ($r_{1/2}=8$ kpc), 
with the variable-beta modelling choosing a near-perfect elliptical host 
($\beta=0.238$). The full extent of the tidal disruption is revealed by 
the model-subtracted image.   
This is the most radio luminous of the `radio-quiet' quasars in our
sample, and its radio morphology is similar to that 
found in several radio-loud quasars with an unresolved core accompanied
by a secondary component $\simeq$ 2 arcsec west of the nucleus 
(Miller, Rawlings \& Saunders 1993; Kukula {\it et al.} 1998a).

\vspace*{0.2in}

\noindent
{\bf0244$+$194} (1E 0244$+$1928)

\vspace*{0.07in}

\noindent
Radio Quiet; ($z$ = 0.176, $\log_{10}(L_{5GHz}/{\rm W Hz^{-1} sr^{-1}})$
$<$ 21.43)

\vspace*{0.07in}

\noindent
The host of this radio-quiet quasar is well described by an 
large elliptical galaxy with $r_{1/2}=19$ kpc.  The variable-beta 
modelling provides a slightly improved fit with a value of $\beta=0.220$.  
The model-subtracted image, shown in figure 1m, is 
very clean and shows no obvious 
signs of any interaction.

\vspace*{0.2in}

\noindent
{\bf0257$+$024} (US 3498)

\vspace*{0.07in}

\noindent
Radio Quiet; ($z$ = 0.115, $\log_{10}(L_{5GHz}/{\rm W Hz^{-1} sr^{-1}})$
= 22.19)

\vspace*{0.07in}

\noindent
The host of this radio-quiet quasar is dominated by a disc component
best-fit scale length of $r_{1/2}=10$ kpc.  The model-subtracted 
image in figure 1n shows a ring of emission at $\approx4$ arcsec radius.  
Also revealed is substantial residual flux in the inner $\approx2$ arcsec 
where the host galaxy is bulge dominated.  This feature can also be 
clearly seen in the luminosity profile shown in figure 2.  As a result of the 
central bulge, the variable-beta model chooses a substantially 
lower value of beta ($0.75$), and offers a 
significantly better $\chi^{2}$ fit than a pure exponential disc.
Nevertheless, this transpires to be the most disc-dominated host galaxy
in the quasar sample observed to date.

\vspace*{0.2in}

\noindent
{\bf0923$+$201} (PG 0923$+$201, Ton 1057)

\vspace*{0.07in}

\noindent
Radio Quiet; ($z$ = 0.190, $\log_{10}(L_{5GHz}/{\rm W Hz^{-1} sr^{-1}})$
$<$ 21.66)

\vspace*{0.07in}

\noindent
This radio-quiet 
quasar is a member of a small group of galaxies.  The two large galaxies 
in figure 1o to the SE and SW are at the same redshift as the quasar 
(Heckman {\it et al}\, 1984), and 0923$+$201 was thought to be 
possibly interacting with these (Hutchings, Janson \& Neff 1989).  However, our 
two-dimensional modelling provides little evidence for any obvious 
interaction.  The host is extremely 
well matched by a standard elliptical template, with the variable-beta 
model also choosing an approximately de Vaucouleurs model ($\beta=0.3$).  
The subtracted image is very clean and symmetrical, but with perhaps a 
suggestion of more residual flux in the direction of the SE companion.

\vspace*{0.2in}

\noindent
{\bf0953$+$415} (K438$-$7, PG 0923$+$415)

\vspace*{0.07in}

\noindent
Radio Quiet; ($z$ = 0.239, $\log_{10}(L_{5GHz}/{\rm W Hz^{-1} sr^{-1}})$
$<$ 21.69) 

\vspace*{0.07in}

\noindent
This radio-quiet quasar is the most heavily nuclear-dominated object 
in the sample.  This can be seen immediately by its stellar-like appearance 
in the greyscale of figure 1p.  The best-fitting host is a medium-sized 
elliptical of scalelength $r_{1/2}=7$ kpc.  
However, due to the faintness of the host relative to the unresolved nuclear 
component, we are unable to exclude a disc host with a high level of 
confidence.  Despite this, we are still able to constrain the relative 
luminosity of the AGN and host components, since both the alternative 
host models yield similar values.  0953$+$415 is another 
object which was imaged with the HST in $V$ by Bahcall {\it et al}
(1997,1995c,1994). From their $V$-band image Bahcall {\it et al.} 
were unable to clearly detect a bright host galaxy centred on the quasar. 
It seems likely that this inconsistency arises primarily from the procedure 
used by Bahcall {\it et al.} to subtract the PSF. 
As can be seen from the luminosity profile shown in figure 2, 
the unresolved AGN component dominates to a radius of $\approx2$ arcsec.  
Normalization of the PSF in a annulus between $1$ and $3$ arcsec 
(as performed by Bahcall {\it et al.} 1997) will therefore lead to a 
substantial over-subtraction. Tests on our own data show that this effect
is more than sufficient to explain the difficulty experienced by 
Bahcall {\it et al.} in detecting the host of this quasar, particularly 
given that this is the most nuclear-dominated object in our sample.
The model-subtracted image reveals a tidal arm feature to the 
SW along with up to four possible companion objects.   
\vspace*{0.2in}

\noindent
{\bf1012$+$008} (PG 1012$+$008)

\vspace*{0.07in}

\noindent
Radio Quiet; ($z$ = 0.185, $\log_{10}(L_{5GHz}/{\rm W Hz^{-1} sr^{-1}})$
= 22.00)

\vspace*{0.07in}

\noindent
This radio-quiet quasar is another dramatically interacting object 
which was also imaged in the $V$--band by Bahcall {\it et al} (1997).  
In the greyscale of figure 1q the quasar is seen together with two companion 
galaxies to the north and east.  The best-fitting host is a large elliptical 
with $r_{1/2}=23$\ kpc, a result firmly supported by the variable-beta 
modelling result ($\beta=0.249$). We note that, using 2-dimensional
modelling, Bachall {\it et al.} (1997) concluded that the host was a disc
galaxy with $r_{1/2} = 10.8$ kpc (after conversion to our adopted
cosmology), however Bahcall {\it et al.}'s best fit to the 1-dimensional
profile of this galaxy is an elliptical with $r_{1/2} = 24.5$ kpc, in
excellent agreement with our result (this perhaps indicates that there is a
problem with the method of 2-dimensional modelling implemented by Bahcall
{\it et al.}) 
From the model-subtracted 
image shown in Panel D it is clear that the eastern companion 
is definitely interacting, with 
the northern companion perhaps also involved.  

\vspace*{0.2in}

\noindent
{\bf1635$+$119} (MC 2)

\vspace*{0.07in}

\noindent
Radio Quiet; ($z$ = 0.146, $\log_{10}(L_{5GHz}/{\rm W Hz^{-1} sr^{-1}})$
= 23.02)

\vspace*{0.07in}

\noindent
The host of this radio-quiet quasar is best-matched by a moderate-sized 
elliptical ($r_{1/2}=6$ kpc) with a disc host being formally excluded. 
The model-subtracted image in figure 1r shows residual luminosity around the core of the 
quasar, unaccounted for by the standard elliptical template.  This 
is confirmed by the variable-beta modelling which shows a statistically 
significant improvement in the quality of fit with a beta parameter of 
$\beta=0.184$. The model-subtracted image also shows numerous companion 
objects in this field.
\vspace*{0.2in}

\noindent
{\bf2344$+$184}

\vspace*{0.07in}

\noindent
Radio Quiet; ($z$ = 0.138, $\log_{10}(L_{5GHz}/{\rm W Hz^{-1} sr^{-1}})$
$<$ 21.18)

\vspace*{0.07in}

\noindent
The modelling of this object presented here must still be regarded as
preliminary. The raw $R$-band 
image presented 
in figure 1s shows this quasar to be residing in a disc galaxy with clearly 
evident spiral arms.  A two dimensional disc 
template of scalelength $12$ kpc can reasonably reproduce the 
total luminosity of the object but is obviously unable to 
cope with the spiral arms. As can clearly be seen from both the 
luminosity profile and model subtracted image, the inner 
$\approx 2$ arcsec are dominated by a bulge with a 
different position angle from the main disc.  
A two component model will therefore 
have to be adopted in order to properly analyse 
this quasar host.

\section{Discussion}

An full statistical analysis of our data is not appropriate until all the
sources in our matched samples have been observed. However, the initial
results presented above suggest the emergence of several potentially 
important trends
which deserve comment at this stage. Furthermore, since the majority 
of RQQs in our
sample have in fact been observed, this interesting subsample of 
objects is already approaching completion and we can draw some fairly
robust conclusions regarding the nature of the radio-quiet quasar
population.

\subsection{Host galaxy luminosities and morphologies}

  The most striking initial result of this study is that our observing
strategy combined with modelling incorporating a high dynamic range PSF
has enabled us not only to easily detect all of the host galaxies
observed to date, but also to determine unambiguously the morphological
type of the host in virtually every case (see tables 2 \& 3, and figures
2 \& 3). 
This represents a major improvement on
previous HST-based quasar-host studies, which in some cases have actually
struggled to detect any host galaxy emission (Bahcall {\it et al.}
1994;1997) leading the authors to conclude that some powerful quasars lie in 
low-luminosity host galaxies (see also McLeod \& Rieke 1995). Our
results to date (which include imaging of several of the quasars also
imaged by Bahcall {\it et al.}) indicate that such conclusions were erroneous.
In contrast, as detailed in table 4 and illustrated in figure 4, we 
confirm the basic result of our ground-based 
$K$-band study that {\it all}
the quasars in our sample lie in host galaxies with luminosities $\ge 
2L^{\star}$ (see the notes on 0953$+$415 in section 3 for a detailed
discussion of the most likely explanation for Bahcall {\it et al.}'s
struggle to detect this host galaxy). 
A similar result with regard to host luminosities
has recently been reported 
from the HST imaging of LBQS 
quasars at $z > 0.4$ by Hooper {\it et al.} (1997), 
(although in this work no attempt was made to identify the morphological 
type of the hosts of these slightly higher-redshift quasars).

   As hoped, these HST images have enabled us to substantially improve on our 
$K$-band study in terms of discerning the morphological type of the hosts. 
All of the radio galaxies and RLQ hosts we have observed to date can be 
unambiguously classified as massive elliptical galaxies, with scalelengths 
in the range 6 kpc $\rightarrow$ 23 kpc, consistent with unification of RGs 
and RLQs via orientation effects. Perhaps more unexpectedly, all except the two
least luminous RQQs are also found to lie in massive ellipticals, placing
on a firm statistical footing the tentative result reported by Taylor
{\it et al.} (1996) and Disney {\it et al.} (1995) 
that luminous RQQs lie in massive ellipticals despite
the fact that the less luminous radio-quiet Seyfert galaxies are
predominantly disc-dominated. Thus, whatever the true physical 
origin of radio loudness, 
it is clearly not simply a consequence of the morphological type of the host
galaxy. A particularly striking feature of this result is the extent to
which a pure $r^{1/4}$ law provides an essentially perfect description
(apart from obvious tidal tails and secondary nuclei) of
the hosts of 16 out of the 19 AGN observed in our sample to date.
 
\subsection{Relation to `normal' massive elliptical galaxies}

\subsubsection{Scalelengths, the Kormendy Relation and axial ratios}

\noindent
The distribution of derived host-galaxy scalelengths is shown in figure 5.
With the exception of the RLQ 2141$+$175, for which accurate scalelength determination
is problematic due to the complexity of the observed interaction, the
half-light scalelengths of all the host galaxies are consistent to within
a factor of three. The homogeneity of these host galaxies is
striking when they are plotted on the $\mu_e - r_e$ projection
of the fundamental plane (see
figure 6) where we find that they describe a Kormendy relation
essentially identical to that displayed by `normal' massive ellipticals
(Schneider, Gunn \& Hoessel 1983; Capaccioli, Caon \& D'Onofrio 1992); a
least-squares fit yields the relationship
$\mu_{1/2} = 3.34_{\pm 0.50}\log_{10} r_{1/2} + 17.95_{\pm 0.53}$.
Thus the basic morphological parameters of these host galaxies appear 
to be indistinguishable from those of normal, 
inactive massive ellipticals.
   
If this conclusion is correct, then the host galaxies should also display a
distribution in axial ratios which is indistinguishable from that
displayed by the normal elliptical galaxy population, which peaks at $b/a
> 0.8$ (Sandage, Freeman \&
Stokes 1970; Ryden 1992). The axial ratios yielded by the model-fitting
are given in table 2, and the resulting host-galaxy axial-ratio distribution is
plotted in figure 7. It is perfectly consistent with the distribution
displayed by normal ellipticals and, with only one object displaying an
axial ratio $b/a < 0.6$, is completely at odds with the recent results of
Hooper {\it et al.} (1997) who reported that {\it most} of the hosts of
bright quasars at $z \simeq 0.4$ have low axial ratios $b/a < 0.6$.
However Hooper {\it et al.} expressed the concern that their result might
reflect high-surface brightness features such as 
bars or tidal tails, rather than the axial ratio of the underlying
stellar population, and our very different result, based on proper
modelling of the underlying host, indicates that this is almost certainly 
the correct explanation for their apparently contradictory conclusion.

\subsubsection{Colours}

One of the reasons we elected to use the F675W filter rather than the
F606W filter used by Bahcall {\it et al.} was the evidence, gleaned from our
deep off-nuclear spectroscopy of quasar hosts (Kukula {\it et al.} 1998b; 
Hughes {\it et al.} 1998),
that most quasar hosts appear to be rather red galaxies with a clear
4000\AA\ break in their spectrum. We are now, for the first
time, in a position
to check whether this is indeed the case by combining our new
$R$-band HST results with the $K$-band results of Taylor {\it et al.}
(1996) to measure the $R-K$ colours of the RGs and quasar hosts. The
results are listed in table 4, and plotted in figure 8, 
where the observed colours are compared
with those predicted from simple $k$-correction of 
stellar populations with ages of 8, 12 and 16
Gyr (Guiderdoni \& Rocca-Volmerange 1987). 
The uncertainty in the $R-K$ colours of the radio galaxies is only
$\simeq 0.15$ mag, and it is striking that these 4 objects track almost
perfectly the $k$-correction derived from the SED of an old (12 Gyr) elliptical 
galaxy. The uncertainties associated with removal of the nuclear
contribution mean that the $R-K$ colours of the quasar hosts are
somewhat more uncertain ($\simeq 0.3 - 0.5$ mag.) but with the present
data it is clear that the hosts of all 3 classes
of powerful AGN have colours which are consistent with each other, and
with that of old passively evolving stellar populations. 
Model independent support for this conclusion can be gleaned by comparing
figure 8 with the $R-K$ versus $z$ plot for objects detected in a $K$-band
survey reaching $K \simeq 17.3$ (Glazebrook {\it et al.} 1995); 
the host galaxy $R-K$ colours shown in figure
8 track well the red envelope displayed by $K$-band selected galaxies at
comparable redshifts.

This is the first clear evidence that the dominant stellar populations in
not only RGs, but also quasar
hosts have ages comparable to the oldest known elliptical galaxies,
and thus must have formed at high redshift ($z > 4$; Dunlop {\it et al.}
1996). It also means
that any substantial star-formation activity associated with the
triggering of AGN activity must either be dust enshrouded, or confined either to the nuclear regions
of the galaxy (in which case it will have been attributed to the
quasar nucleus in our modelling procedure) or to obvious tidal tails,
which we have excised prior to modelling the host galaxy.

\subsubsection{Interactions}

As can be seen from Panel D in figures 1a$-$1s, removal of the 
axisymmetric model for the underlying host galaxy makes it relatively 
easy to identify morphological peculiarities such as excess flux, tidal
tails, interacting companions and secondary nuclei, often at considerably
lower surface brightness levels than the more obvious morphological
distortions masked out prior to host-galaxy modelling.
A full statistical analysis of the prevalence and strength of such
features is deferred until completion of the sample, but here we simply
note that, despite our basic conclusion that the host galaxies 
are relatively passive massive ellipticals, 3 out of the 4 RGs, 5 out of
the 6 RLQs and 6 out of the 9 RQQs observed to date show one or other of 
the morphological peculiarities mentioned above.

This tally serves to emphasize that most of these AGN may have been
triggered into action by the interaction of their (perhaps previously
completely passive) host galaxy with a companion object. However,
we add the cautionary note that the true significance of such apparently 
impressive interaction statistics (14 out of 19 AGN) 
can only really be judged against the results
of a comparably-detailed investigation of the morphologies of `inactive'
massive ellipticals.

\subsection{The AGN -- host connection}

The absolute magnitudes $M_R$ of the fitted nuclear components
are given in table 4, and the resulting distribution is 
presented in figure 9. The nuclear absolute magnitudes are then plotted
against the host absolute magnitudes in figure 10. Only a weak
correlation ($p = 0.376$, using the Spearman Rank correlation test) 
is seen, consistent with our previous finding at $K$.
However, the fact that the hosts of the two faintest quasars in our sample  
contain a significant disc component suggests that a more significant
correlation might be revealed if only bulge luminosity is considered for
comparison with nuclear power. This possibility is explored below in the
context of recent studies of nearby galaxies which suggest that black
hole mass does indeed depend primarily 
on bulge mass rather than total galaxy mass.
 
\subsubsection{The black hole -- spheroid connection}
 
Our unambiguous finding that all the quasars in our sample 
with $M_R < -23.5$ lie in massive elliptical galaxies, irrespective of 
radio power, clearly refutes the long-standing 
hypothesis that, like the majority of Seyferts, radio-quiet quasars
lie in predominantly disc galaxies.
However, if it is accepted that all quasars (radio-loud and quiet) 
result from accretion of material 
onto a supermassive black hole, then our result can
be seen as a natural consequence of the black-hole/spheroid mass
correlation recently derived in nearby galaxies by Magorrian {\it et al.}
(1998), as we now briefly explain.

Magorrian {\it et al.} find that the available kinematic data 
on nearby galaxies 
are consistent with the relation $m_{bh} = 0.006 m_{sph}$ where $m_{sph}$
is the mass of the hot stellar component ({\it i.e.} the spheroidal
bulge), strengthening the previous conclusion of Kormendy \& Richstone
(1995). 
They also show that $m_{sph}$ can be estimated from the luminosity of the
spheroidal component using a mass:light ratio which is proportional to 
$M^{0.18}$ (their equation (10)), broadly consistent with the
fundamental-plane correlation predicted using the virial theorem ({\it
e.g.} Bender, Burstein \& Faber 1992). 

Since we now possess the first reliable determinations
of the spheroid luminosity for the hosts of a significant sample of AGN,
we have explored the result of applying these two relations 
to estimate (albeit rather crudely)
the spheroidal mass $M_{sph}$ of each host, 
and hence the {\it expected}  mass of the black hole at the centre of each 
galaxy. The results of this calculation are listed in table 5, columns 2
and 3.
We have then proceeded to calculate the Eddington luminosity, Eddington
temperature, and hence Eddington absolute magnitude $M_R$ for the putative 
black hole at the centre of each host galaxy, and the results of this
calculation are given in column 6 of table 5, and 
compared with the observed 
nuclear absolute magnitude of each quasar ({\it i.e.} 
after host galaxy removal)
in figure 11.

In our view this is a surprisingly
successful calculation. As can be seen from figure 11, the maximum
luminosity produced by any quasar is comparable with the predicted
Eddington limit, while the majority appear to be radiating at a few per
cent of the Eddington luminosity.
Moreover, despite the large number of sources of potential scatter, the 
correlation between observed $M_R$
and predicted Eddington $M_R$ is stronger ($p = 0.164$) 
than that between raw host-galaxy
luminosity and nuclear-luminosity (and becomes marginally 
significant ($p = 0.07$)
if 0953$+$415 is excluded from the analysis), and both the radio-quiet and
radio-loud quasars display a similar relation. 
This suggests that in both
classes of quasar the {\it optical} luminosity arises from a similar process of
accretion onto a massive black hole, and that RLQs and RQQs of comparable
absolute magntiude $M_R < -23.5$ 
are powered by black holes of comparable mass $M > 
3 \times 10^{9} {\rm M_{\odot}}$. 

The results of this calculation can be regarded as providing independent
evidence that the relation of Magorrian {\it et al.} still applies for large galaxy
and black hole masses. To reiterate, this relation 
would predict that such massive black holes can only be housed
in galaxies with a spheroid mass of $> 5 \times 10^{11} {\rm M_{\odot}}$
which is equivalent to an absolute magnitude of $M_R < -23$, which is
exactly what we find. 
Galaxies with such massive spheroidal components must inevitably be
classed as giant ellipticals, in which case it could be regarded as a
(now successful) prediction of the black-hole/spheroid mass relation 
that all luminous quasars must reside in massive elliptical
galaxies more luminous than $\simeq 2 L^{\star}$. 

\subsubsection{The black hole -- radio power connection}

Finally we combine our estimates of black hole mass derived from 
host-galaxy spheroid mass (table 5)  
with the radio data available to us for each object to investigate how
our results compare with the radio-luminosity--black-hole mass
correlations recently derived for low-redshift galaxies by Franceschini,
Vercellone \& Fabian (1998). 

Franceschini {\it et al.} found a remarkably tight relationship between 
black hole mass and both total and nuclear radio centimetric luminosity, with a
very steep dependence of the radio power on the mass of the black hole
$m_{bh}$ ($P_{5GHz}^{tot} \propto m_{bh}^{2.7}$). 
In figure 12a we have replotted
their data which demonstrate the relation between total 5GHz radio
luminosity and black hole mass in nearby galaxies, along with
their best-fit straight-line relation, highlighting 
the location of the Milky Way and M87. We have 
then added the relevant datapoints (or upper limits) for the
AGN studied in this paper. Whereas
Franceschini {\it et al.} were able to determine $m_{bh}$ directly from
high-resolution spectroscopy, the value of $m_{bh}$ plotted for each of
our AGN has of
course had to been inferred from the mass of the spheroidal component of
its host galaxy. However, the location of our AGN on this diagram,
particularly the radio-loud AGN, strongly suggests that our host-galaxy-based 
black hole mass estimates are reasonable, despite the large
uncertainties involved in extrapolating from spheroid luminosity.

To put this another way, one can derive two completely independent estimates
the mass of the black hole at the heart of each of our AGN 
using either i) the luminosity of
the host spheroid and the relations given by Magorrian {\it et al.} as
described above, or ii) the $P_{5GHz}$:$m_{bh}$ regression line of 
Franceschini {\it et al.} shown in figure 12a, and for the radio loud AGN
in our sample these values agree to within a factor of typically 
2 (r.m.s. - see columns 3
and 4 of table 5).

Since the RGs and RLQs appear so consistent with the $P_{5GHz}:m_{bh}$
relation, it is inevitable that the RQQs should lie below it, but a
striking feature of figure 12a is that (albeit that a number of upper
limits are involved), there does still appear to be a correlation between
radio-luminosity and host-galaxy-derived $m_{bh}$ within the RQQ subsample.
This suggests that the radio luminosities of the RQQs might also be
linked to those of lower mass objects via a simple relation. Since the
study of Kukula {\it et al.} (1998) has shown that the radio emission
from these RQQs, where detectable, arises entirely from a compact
core-like component, we have therefore investigated whether the radio
properties of RQQs can be linked to those of nearby galaxies if only
compact radio emission is included.

This is explored in figure 12b. Franceschini {\it et al.} also found a
strong correlation between core 5GHz radio luminosity and black hole
mass, and so we have again reproduced their data for nearby galaxies,
this time plotting $P_{5GHz}^{core}$ versus $m_{bh}$, along with the
relation $P \propto m_{bh}^{2.2}$ expected for simple advection-dominated
accretion models (Fabian \& Rees 1995), or indeed for any 
model in which the emission is mostly dependent of the emitting area
available around a black hole. We have replotted all the RQQ data points
shown in Figure 12a, but this time have only plotted datapoints for the
cores of the lobe-dominated radio-loud AGN (to minimize the impact of
beaming; Giovannini {\it et al.} 1988;
Clarke {\it et al.} 1992;
Lister \& Gower 1994). We note that in this case the RQQs
apparently form a natural extension of the local galaxy sample, as do the
least luminous cores of the radio-loud AGN, and that 
the plotted relation provides an excellent description of the data. 
A number of 
points are worthy of comment. First, unlike Franceschini {\it et al.} we have
attributed all the radio emission from M31 (after removal of very extended 
radio emission linked to star-formation) to the core because the AGN 
contribution in M31 would in 
fact remain unresolved at the distance of most of the objects in even the 
local sample; this appears to
the reason that we find the core radio data to be linked to $m_{bh}$
through a flatter relation $P_{5GHz}^{core} \propto m_{bh}^{2.2}$ than did 
Franceschini {\it et al.}. Second, it is noteworthy that M87 moves from
the `radio-loud' relation in figure 12a, to the `radio-quiet' relation in
figure 12b once its extended radio emision is removed, as do several of
the radio-loud AGN. Third, as is clear from the figure,
for several of the RQQs we currently only possess upper limits on radio 
luminosity; if $P_{5GHz}^{core}$ and $m_{sph}$ really are both good
indicators of the black hole mass in these objects, then we would predict
that the radio emission from these RQQs should be detected by radio
observations reaching only an order of magnitude below the current limits.

Because it is undoubtedly possible that $P_{5GHz}^{core}$ is a better
indicator of black hole mass than is $m_{sph}$, we have used the relation
shown in Figure 12b to obtain an independent prediction of $m_{bh}$ for
each quasar, and the results are presented in column 5 of table 5. In
figure 13 we plot $m_{bh}$ derived from $P_{5GHz}^{core}$ against 
$m_{bh}$ derived from host galaxy luminosity. These {\it completely independent 
estimates} of $m_{bh}$ are clearly extremely 
well correlated ($p = 0.003$), and for individual
sources the discrepancy is no greater than a factor of 4, and is
frequently smaller. However, use of the $P_{5GHz}^{core} - m{bh}$
relation shown in figure 12b, does 
tend to yield smaller black hole masses for the RQQs than the
use of $m_{sph}$.

\subsection{The origin of radio loudness}

Despite the good agreement between the black hole masses of the
radio-loud AGN as estimated from $P_{5GHz}^{total}$ and host galaxy
luminosity, it seems unlikely that the relation shown in Figure 12a can
be a fair indicator of the black hole masses in the RQQs because then
the more luminous RQQs such as 0054$+$144 would appear to have an optical
luminosity an order of magnitude greater than the Eddington luminosity.

This leaves either $P_{5GHz}^{core}$ or host-galaxy luminosity as
potential alternative estimators of black-hole mass, and while figure 13
demonstrates that both quantities lead to reasonably consistent values
for $m_{bh}$, the implications for the physical origin of radio loudness 
 depend rather crucially on which of these is the more
reliable predictor of $m_{bh}$. The reason for this is that, for the
radio-loud objects, comparison of columns 3, 4 and 5 in table 5 shows that
essentially all routes of black-hole mass estimation lead to values of
$m_{bh} > 10^{10} M_{\odot}$. However, for the RQQs, use of
$P_{5GHz}^{core}$ to estimate $m_{bh}$ produces values typically a factor 
of two smaller than inferred from host-galaxy luminosity, and implies
that {\it no} RQQ in the sample has a black hole more massive than
$m_{bh} = 10^{10} M_{\odot}$. There is already a suggestion in figure 11
that, despite the comparable observed absolute magnitudes of the RQQs and
RLQs, the predicted Eddington luminosities of the RQQs are somewhat
smaller. If the values of $m_{bh}$ derived from $P_{5GHz}^{core}$
are use to predict Eddington $M_R$, the difference becomes even more
stark (as shown in figure 14), and shows that in selecting
optically matched samples of RQQs and RLQs we may in fact have selected
lower mass ($\simeq 10^{9} M_{\odot}$) black holes radiating close to the
Eddington limit for comparison with higher mass ($\simeq 10^{10} M_{\odot}$)
black holes radiating (in the optical) at around 10\% of their Eddington
luminosity.

In summary, our results to date can be interpreted in two distinct ways
depending on whether $P_{5GHz}^{core}$ or host-galaxy luminosity is the 
the more reliable predictor of black hole mass. If the former, then the
difference between radio-loud and radio-quiet AGN may simply be that FRII
radio sources require black holes of mass $m_{bh} > 10^{10} M_{\odot}$.
If the latter, then at least some of the RQQs in our sample would also
appear to be powered by black holes with $m_{bh} > 10^{10} M_{\odot}$,
and some other explanation ({\it e.g.} black hole angular momentum) 
would be required to explain why 
black holes of comparable mass can produce radio sources which differ by
two orders of magnitude in radio power, despite the fact that both are
radiating in the optical with comparable efficiency.

Completion of our sample, coupled with proposed deeper radio observations
of the as yet undetected RQQs should assist in clarifying this issue.

\section{Conclusions}

In this paper we 
have presented and analyzed the deep WFPC2 R675W images of the 19 objects
(RGs, RLQs and RQQs) observed with the HST during the first year of our
comparative HST imaging study of the host galaxies of luminous AGN.
The results indicate that this carefully controlled 
study will, as hoped, be able to identify
unambiguously the morphological type of the host galaxy of every
AGN in our final 33-source sample. 
From the images analyzed in this paper we find
that the underlying hosts of {\it all three} 
classes of luminous AGN are
massive elliptical galaxies, with scalelengths $\simeq 10$ kpc.
Since the RQQ sub-sample is already close to completion we can therefore
for the first time state with some confidence
that essentially all RQQs brighter than $M_R = -24$ 
reside in massive ellipticals, a
a result which removes the possibility that radio `loudness' is directly 
linked to host galaxy morphology.

The existence of comparably detailed modelling of deep $K$-band images of
all the sources in our sample (Dunlop {\it et al.} 1993; Taylor {\it et
al.} 1996) has allowed us to use our new $R$-band results to derive the
first reliable optical-infrared colours for the hosts of a substantial
sample of AGN. The preliminary results indicate that the hosts of all
three classes of AGN have underlying stellar populations with ages 
comparable to that observed in normal, old ellipticals at similar
redshift, despite the evidence for superimposed acivity 
arising from interactions/mergers. We also find that the 
distribution of host-galaxy axial
ratios is consistent with that displayed by the normal elliptical
galaxy population.

The RG and RLQ sub-samples have been designed to be matched in terms of
radio luminosity, radio spectral index, and redshift. 
Comparison of the host galaxy
properties of the (still small) RG and RLQ sub-sub-samples observed to date
indicates that their hosts are indistinguishable in terms of
morphological type,
luminosity, axial ratio, and $R-K$ colour, consistent with unification
via orientation.

The RQQ and RLQ samples were designed to be matched in terms of optical
luminosity and redshift, and the image analysis presented here indicates
that, with the exception of two low-luminosity interlopers in the RQQ
sample (0257$+$024 and 2344$+$184, which should really be reclassified as
Seyferts) the distribution of nuclear absolute magnitudes of the RQQ and
RLQ sub-samples do indeed appear to be well matched. Interestingly, therefore,
with the exception of 0257$+$024 and 2344$+$184, we find the host
galaxies of the RQQs to also be massive elliptical galaxies, albeit
approximately 0.5 mag fainter on average than their radio-loud
counterparts. With the present sub-samples this difference is not
significant, but when the black hole mass in these quasars is inferred
from the luminosity of their host spheroid, it implies that the typical
black hole mass for the RQQs in our sample is $\simeq 7 \times 10^{9}
{\rm M_{\odot}}$, while for the RLQs it is $\simeq 15 \times 10^{9}
{\rm M_{\odot}}$, and that several of 
the luminous RQQs we have selected are radiating close to their Eddington
limit (whereas very few of the RLQs appear to be radiating at more than
$\simeq 10$\% of predicted Eddington luminosity).

Finally we find that the black hole mass estimates obtained 
from the luminosity
of the host spheroid are in good agreement with independent estimates
based on extrapolation of the relation between black-hole mass and 
core radio luminosity established for nearby galaxies, 
and it seems hard to escape the conclusion that
all the radio-loud AGN in our sample have black hole masses in excess of
$10^{10} {\rm
M_{\odot}}$. However,
if the latter relation is adopted as the
most reliable estimator we would conclude that our RQQ sample is even
more biased towards objects emitting close to their Eddington limit, and
contains no object powered by a black hole as massive as $10^{10} {\rm
M_{\odot}}$.
If this is correct, then the physical origin of radio loudness( $P_{5GHz}
> 10^{24} {\rm W Hz^{-1} sr^{-1}}$) may simply be
the presence of a black hole more massive than 
$10^{10} {\rm
M_{\odot}}$, and this relatively clean result may have been previously concealed
from us by the selection effects involved in striving to produce
a bright RQQ
sample for comparison with a (radio-selected) RLQ sample, coupled with
substantial scatter in the $m_{bh}$:host-luminosity relation.
Alternatively, if our values of $m_{bh}$ derived from host-galaxy
spheroid luminosity are more reliable, then at least some of the RQQs in
our sample would have black hole masses $m_{bh} > 10^{10} {\rm
M_{\odot}}$, comparable to RLQs, and some explanation other than simply
black hole mass would still be required to account for the fact that the
RLQs are two orders of magnitude more luminous at radio wavelengths.
Completion of our HST study, coupled with deeper radio observations of
the as-yet-undetected RQQs should help to distinguish between these two
alternative scenarios.

\section*{Acknowledgements}
This work was supported by PPARC through the award of a studentship to RJM, and
the award of PDRAs to MJK and DHH. Support was also provided 
by NASA through GO program grant 6776 from the
Space Telescope Science Institute, which is operated by the Association of
Universities for Research in Astronomy Inc., under NASA contract NAS
5-26555. 

\section*{References}

\noindent
Akujor C.E., Spencer R.E., Zhang F.J., Davis R.J., Browne I.W.A., Fanti C., 1991, MNRAS, 250, 215\\
Antonucci R.J.J., 1985, ApJ, 59, 499\\
Bahcall J.N., Kirhakos S., Schneider D.P., 1994, ApJ, 435, L11\\
Bahcall J.N., Kirhakos S., Schneider D.P., 1995a, ApJ, 447, L1\\
Bahcall J.N., Kirhakos S., Schneider D.P., 1995b, ApJ, 450, 486\\
Bahcall J.N., Kirhakos S., Schneider D.P., 1996, ApJ, 457, 557\\
Bahcall J.N., Kirhakos S., Schneider D.P., 1997, ApJ, 479, 642\\
Barthel P.D., 1989, ApJ, 336,606\\
Baum, S.A., Heckman, T.M., Van Breugel, W., 1992, ApJ, 389, 208\\
Capaccioli M., Caon N., D'Onofrio M., 1992, MNRAS, 259, 323\\
De Koff S., et al., 1996, ApJS, 107, 621\\
Disney M.J. et al., 1995, Nature, 376, 150\\
Dunlop J.S., 1997,
In: {\em Observational Cosmology with the New Radio Surveys}, p.157
eds Bremer, M., et al., Kluwer (astro-ph/9704294)\\
Dunlop J.S., 1998, In: {\em `The Most Distant Radio Galaxies'}, 
KNAW Colloquium Amsterdam, eds. 
Rottgering, H.J.A., Best, P., Lehnert, M.D., Kluwer, in press
(astro-ph/9801114)\\
Dunlop J.S., Taylor G.L., Hughes D.H., Robson E.I., 1993, MNRAS, 264, 455\\
Ellingson E., Yee H.K.C., Green R.F., 1991, ApJ, 371, 41\\
Fabian A.C., Rees M.J., 1995, MNRAS, 277, L55\\
Feigelson E.D., Isobe T., Kembhavi A., 1984, AJ, 89, 1464\\
Franceschini A., Vercellone S., Fabian A.C., 1998, MNRAS, 297, 817\\
Giovannini, G., Feretti, L., Gregorini, L., Parma, P., 1988, A\&A, 199,
73\\
Guiderdoni B., Rocca-Volmerange B., 1987, A\&A, 186, 1\\
Glazebrook K., Peacock J.A., Miller L., Collins C.A., 1995, MNRAS, 275,
169\\
Gower A.C., Hutchings J.B., 1984a, PASP, 96, 19\\
Gower A.C., Hutchings J.B., 1984b, AJ, 89, 1658\\
Haehnelt M.G., Rees M.J., 1993, MNRAS, 263, 168\\
Heckman T.M. et al, 1984, ApJ, 89, 958\\
Heckman T.M. et al, 1986, ApJ, 311, 526\\
Hooper E.J., Impey C.D., Foltz C.B., 1997, ApJ, 480, L95\\
Hughes D.H., Kukula M.J., Dunlop J.S., Boroson T.B., 1998, MNRAS,
submitted\\
Hutchings J.B., Neff S.G., 1992, AJ, 104, 1\\
Hutchings J.B., 1995, Nature news \& Views, 376, 118\\
Hutchings J.B., Morris S.C., 1995, AJ, 109, 1541\\
Hutchings J.B., Johnson I., Pyke R., 1988, ApJS, 66, 361\\
Hutchings J.B., Janson T., Neff S.G., 1989, ApJ, 342, 660\\
Hutchings J.B., Holtzman J., Sparks W.B., Morris S.C., Hanisch R.J., Mo,
J., 1994, ApJ, 429, L1\\
Kormendy J., Richstone D., 1995, ARA\&A, 33, 581\\
Kukula M.J., Dunlop J.S., Hughes D.H., Rawlings S., 1998a, MNRAS, 297,
366\\
Kukula et al, 1997, in: Quasar Hosts, proc. ESO/IAC conference, eds.
Clements, D.L. \& P\'{e}rez-Fournon, I., Springer-Verlag:Berlin, p. 177\\
Leahy J.P., Pooley G.G., Riley J.M., 1986, MNRAS, 222, 753\\
Lin H., Kirshner P.P., Schectman S.A., Landy S.D., Oemler A., Tucker
D.L., Schechter P.L., 1996, ApJ, 464, 60\\
Lister, M.L., Gower, A.C., 1994, AJ, 108, 821\\
McCarthy P.J., van Breugel W., Kapahi V.K., 1991, ApJ, 371, 478\\
McLeod K.K., Rieke G.H., 1995, ApJ, 454, L77\\
McLure R.J., et al, 1998b, MNRAS, in preparation\\
MacKenty J.W., 1990, ApJS, 72, 231\\
Magorrian J., et al., 1998, AJ, 115, 2285\\
Miley G.K., Hartsuijker A.P., 1978, A\&AS, 34, 129\\
Miller P., Rawlings S., Saunders R., 1993, MNRAS, 263, 425\\
Peacock J.A., 1987, In: {\it Astrophysics Jets and Their Engines}, p.185,
ed. Kundt W., Reidel.\\
Romney J. et al., 1984, A\&A, 135, 289\\
Ryden, S., 1992, ApJ, 396, 445\\
Sandage A.R., Freeman K.C., Stokes N.R., 1970, ApJ, 160, 831\\
Schneider D.P., Gunn J.E., Hoessel J.G., 1983, ApJ, 268, 476\\
Silk J., Rees M., 1998, in press, (astro-ph/9801013)\\
Small T.A., Blandford R.D., 1992, MNRAS, 259, 725\\
Smith E.P., Heckman T.M., 1990, ApJ, 348, 38\\
Smith E.P., Heckman T.M., Bothun G.D., Romanishin W., Balick B., 1986, ApJ, 306, 64\\
Stockton A., Farnham T., 1991, ApJ, 371, 525\\
Taylor G.T., Dunlop J.S., Hughes D.H., Robson E.I., 1996, MNRAS, 283, 930\\ 
Trauger et al., 1994, ApJ, 435, L3\\
Turnshek, D.A., Bohlin, R.C., Williamson, R., Lupie, O., Koornneef,
  J., \& Morgan D. 1990, AJ, 99, 1243\\
Urry C.M., Padovani P., 1995, PASP, 107, 803\\
van Breugel W., Miley G., Heckman T., 1984, AJ, 89, 5\\
V\'{e}ron-Cetty M.P., Woltjer L., 1990, A\&A, 236, 69\\
Young S. et al, 1998, MNRAS, 294, 478\\
Zirbel, E.L., Baum, S.A., 1995, ApJ, 448, 521\\

\newpage

\section*{Figure Captions}

\begin{figure}
\caption{The images, two-dimensional model fits, and model-subtracted
residual images. 
A grey-scale/contour image of the final reduced 
F675W $R$-band image of each AGN 
is shown in the top
left panel (panel A) of each figure 1a$-$1s, 
which shows a region 12.5 $\times$ 12.5 arcsec
centred on the target source. The surface brightness of the lowest contour 
level is indicated in the top-right corner of the panel with the greyscale 
designed to highlight structure close to this limit.  Higher surface brightness 
contours are spaced at intervals of 0.5 mag. arcsec$^{-2}$, and have been
superimposed to emphasize brighter structure in the centre
of the galaxy/quasar.  Panel B in each figure shows the best-fitting 
two-dimensional model, complete with unresolved nuclear component (after
convolution with the empirical PSF) contoured in an identical manner to 
to panel A. 
Panel C shows the best-fitting host galaxy as it would appear if 
the nuclear component were absent, while panel D is the residual image 
which results from subtraction of the full two-dimensional model (in
panel B) from the raw $R$-band image (in panel A), in order to highlight
the presence of morphological peculiarities such as tidal tails,
interacting companion galaxies, or secondary nuclei.
All panels are displayed using the same greyscale.
a) The Radio Galaxy 0345$+$337 (3c93.1);
b) The Radio Galaxy 0917$+$459 (3c219);
c) The Radio Galaxy 0958$+$291 (3c234.0);
d) The Radio Galaxy 2141$+$279 (3c436);
e) The Radio-Loud Quasar 0137$+$012;
f) The Radio-Loud Quasar 0736$+$017;
g) The Radio-Loud Quasar 1004$+$130;
h) The Radio-Loud Quasar 2141$+$175;
i) The Radio-Loud Quasar 2247$+$140;
j) The Radio-Loud Quasar 2349$-$014;
k) The Radio-Quiet Quasar 0054$+$144;
l) The Radio-Quiet Quasar 0157$+$001;
m) The Radio-Quiet Quasar 0244$+$194;
n) The Radio-Quiet Quasar 0257$+$024;
o) The Radio-Quiet Quasar 0923$+$201;
p) The Radio-Quiet Quasar 0953$+$415;
q) The Radio-Quiet Quasar 1012$+$008;
r) The Radio-Quiet Quasar 1635$+$119;
s) The Radio-Quiet Quasar 2344$+$184}
\end{figure}

\begin{figure}
\caption{A comparison of the azimuthally-averaged luminosity profiles
derived from the images with those produced by the two-dimensional
modelling. Each plot shows the azimuthally-averaged
image data (open circles), the azimuthally-averaged best-fit model 
after convolution with the PSF (solid line) and the azimuthally-averaged best-fit 
unresolved nuclear component after convolution with the PSF (dashed line).}
\end{figure}

\begin{figure}
\caption{The distribution of $\beta$ values which results from fitting
luminosity profiles of the form $exp(-r^{\beta})$ to the host galaxies
and allowing $\beta$ to vary as a free parameter in the model fitting.
As can be seen from the histogram, 16 of the 19 AGN have host galaxies 
which follow near-perfect de Vaucouleurs profiles.}
\end{figure}

\begin{figure}
\caption{The distribution of (integrated) $R$-band absolute magnitudes ($M_R$)
displayed by the best-fitting host galaxies. The adopted fiducial
integrated absolute magnitude $M_R^{\star}$ corresponding to an $L^{\star}$ galaxy is
$M_R^{\star} = -22.2$, which is derived from the most recent determination 
of $M_R^{\star}$ by Lin {\it et al.} (1996) ($M_R^{\star} = -21.8$)
after correcting to an integrated magnitude. All the host galaxies have
$L > 2L^{\star}$.}
\end{figure}

\begin{figure}
\caption{The distribution of scalelengths displayed by 
the best-fitting host galaxies. All scalelengths
are presented as half-light radii ($r_{1/2}$) to facilitate ease of 
comparison between the sizes of disc and elliptical hosts.}
\end{figure}

\begin{figure}
\caption{The Kormendy surface-brightness/scalelength 
relation ($\mu_{1/2}$ v $r_{1/2}$) displayed by the host galaxies of the RGs
(crosses), RLQs (open circles) and RQQs (filled circles).
Also shown on the plot are the best-fitting relation described in the text
(solid line; slope $\simeq$ 3), a line indicating the locus of constant
galaxy luminosity (dashed line; slope $=$ 5), and the dividing line 
between normal ellipticals and brightest cluster members as determined by
Capaccioli {\it et al.} (1992) (dot-dash line). }
\end{figure}

\begin{figure}
\caption{The distribution of axial ratios displayed by the model host
galaxies. The distribution is consistent with that 
displayed by the normal elliptical galaxy population, which peaks at $b/a
> 0.8$ (Sandage, Freeman \&
Stokes 1970; Ryden 1992).}
\end{figure}

\begin{figure}
\caption{The apparent $R-K$ colours of the hosts (RGs = crosses, RLQs =
open circles, RQQs = filled circles) plotted against
redshift, compared with the colours predicted 
from simple $k$-correction of 
stellar populations with ages of 8, 12 and 16
Gyr (Guiderdoni \& Rocca-Volmerange 1987). 
The uncertainty in the $R-K$ colours of the radio galaxies is only
$\simeq 0.15$ mag, and it is striking that these 4 objects track almost
perfectly the $k$-correction derived from the SED of an old (12 Gyr) elliptical 
galaxy. The uncertainties associated with removal of the nuclear
contribution mean that the $R-K$ colours of the quasar hosts are
somewhat more uncertain ($\simeq 0.3 - 0.5$ mag.) but with the present
data it is clear that the hosts of all 3 classes
of powerful AGN have colours which are consistent with each other, and
with that of mature stellar populations.}
\end{figure}

\begin{figure}
\caption{The distribution of $R$-band absolute magnitudes ($M_R$)
displayed by the best-fitting unresolved nuclear components.}
\end{figure}

\begin{figure}
\caption{Absolute host galaxy magnitude plotted against absolute nuclear
magnitude at $R$ (RLQs =
open circles, RQQs = filled circles). The correlation is not statistically 
significant ($p = 0.376$), but
is consistent with the existence of a minimum host galaxy luminosity for
the production of a luminous quasar.
One of the four radio galaxies lies outside this diagram due to 
the essentially zero luminosity of its nuclear component.}
\end{figure}

\begin{figure}
\caption{The observed absolute magnitude $M_R$ of the nuclear component
in each quasar plotted against the absolute magnitude which is predicted 
by assuming that each quasar contains a black hole of mass $m_{bh} = 0.006
m_{spheroid}$, and that the black hole is emitting at the Eddington
luminosity(RLQs =
open circles, RQQs = filled circles). The 
black hole mass has been calculated from the host 
galaxy bulge luminosity, assuming the mass to light ratio and 
$m_{spheroid}-m_{BH}$ 
correlation given in Magorrian {\it et al.} (1998). 
The solid line shows where the quasars should lie if they were  all 
radiating at their respective Eddington luminosities, while the dashed
line indicates $10\%$ of predicted Eddingtion luminosity, and the dotted line
indicates $1\%$ of predicted Eddington luminosity. As in previous figures,
the radio-loud quasars are indicated by open circles, while the radio
quiet quasars are indicated by filled circles. The nuclear components of
the radio galaxies are not plotted because all the evidence suggests they
are substantially obscured by dust.}
\end{figure}

\begin{figure}
\caption{a: Total radio luminosity $P_{5GHz}^{total}$ 
versus black-hole mass showing the data on low-redshift `normal' galaxies 
from Franceschini {\it et al.} (1998), and the AGN discussed in this paper 
(RGs = crosses, RLQs = open circles, RQQs = filled circles). The solid
line is simply the best-fitting relation to the nearby galaxy data given by
Franceschini {\it et al.} - $ log (P_{5GHz}^{total}) = 2.73 log (m_{bh}) - 
2.87$. For the nearby galaxies $m_{bh}$ has been estimated directly from
stellar dynamics, while for the AGN $m_{bh}$ has been estimated from
host-galaxy spheroid luminosity using the relations derived by Magorrian
{\it et al.} (1998) 
;b: Core radio luminosity $P_{5GHz}^{core}$ 
versus black-hole mass showing the data on low-redshift `normal' galaxies 
from Franceschini {\it et al.} (1998), and the AGN discussed in this paper 
(RGs = crosses, RLQs = open circles, RQQs = filled circles) for which a
core radio flux was available in the literature with little evidence for
a signficant beamed component. The solid line is the
relation $P \propto m_{bh}^{2.2}$ expected for simple advection-dominated
accretion models (Fabian \& Rees (1995); or indeed for any 
model in which the emission is mostly dependent of the emitting area
available around a black hole) normalized to the Milky Way. 
Unlike Franceschini {\it et al.} we have
attributed all the radio emission from M31 (after removal of very extended 
radio emission linked to star-formation) to the core because the AGN 
contribution in M31 would in 
fact be remain unresolved at the distance of most of the objects in even the 
local sample; this appears to
the reason that we find the core radio data to be linked to $m_{bh}$
through a flatter relation $P_{5GHz}^{core} \propto m_{bh}^{2.2}$ than did 
Franceschini {\it et al.}.}
\end{figure}

\begin{figure}
\caption{Black hole mass ($m_{bh}^c$ - as given in column 5 of table 5)
as estimated from core radio flux $P_{5GHz}^{core}$ plotted against
black hole mass ($m_{bh}^a$ - as given in column 3 of table 5)
estimated from host galaxy luminosity. These two completely independent
estimators of black hole mass are highly correlated ($p = 0.003$), and
for individual objects agree to no worse than a factor of 4}
\end{figure}

\begin{figure}
\caption{As figure 11, but this time with Eddington $M_R$ derived using 
the values of $m_{bh}$ estimated from $P_{5GHz}^{core}$.}
\end{figure}

\newpage

\begin{table}

\begin{tabular}{llcl}
\hline
Object & $HST$ Archive & Type & Observing \\ 
       & designation   &      & date\\ \hline
GRW $+$70D5824 & PSF-STAR & STAR & Aug 07 1997 \\
0958$+$291 & 3C234.0 & RG & Jun 11 1997 \\
0345$+$337 & 3C93.1 & RG & Feb 16 1998 \\
0917$+$459 & 3C219 & RG & Mar 25 1998 \\
2141$+$279 & 3C436 & RG & Apr 27 1998 \\
2247$+$140 & PKS2247$+$14 & RLQ & Jun 25 1997 \\
2141$+$175 & OX169 & RLQ & Jul 01 1997 \\
0137$+$012 & PHL1093 & RLQ & Jul 04 1997 \\
2349$-$014 & PKS2349$-$01 & RLQ & Jul 05 1997 \\
1004$+$130 & PKS1004$+$13 & RLQ & Nov 28 1997 \\
0736$+$017 & PKS0736$+$01 & RLQ & Feb 06 1998 \\
0953$+$415 & PG0953$+$415 & RQQ & Jun 02 1997 \\
0054$+$144 & PHL909 & RQQ & Jun 27 1997 \\
2344$+$184 & 2344$+$184 & RQQ & Jun 28 1997 \\
0244$+$194 & 0244$+$194 & RQQ & Jun 29 1997 \\
0157$+$001 & 0157$+$001 & RQQ & Jun 30 1997 \\
0257$+$024 & US3498 & RQQ & Jul 04 1997 \\
1635$+$119 & MC1635$+$119 & RQQ & Feb 19 1998 \\
0953$+$201 & PG0923$+$201 & RQQ & Mar 12 1998 \\
1012$+$008 & PG1012$+$00 & RQQ & Mar 12 1998 \\ \hline
\end{tabular}
\caption{The AGN for which images and the results of image modelling are
presented in this paper, along with HST observing dates and archive
designations.}
\end{table}

\clearpage

\begin{table}
\begin{tabular}{clrccccccr}
\tableline
\tableline
Source & Host& $\Delta\chi^{2}$ & $r_{1/2}$/kpc  &$\mu_{1/2}$&$R_{host}$
& $R_{nuc}$  & $L_{nuc}/L_{host}$ & $b/a$ & PA/$^{\circ}$ \\
\tableline
{\bf RG}&&&&&&&&&\\
0345$+$337&Elliptical&2349&11.0&23.1&18.0&21.1&0.06&0.70&99\\

0917$+$459&Elliptical&33245&19.1&23.0&16.1&19.4&0.05&0.76&36\\

0958$+$291&Elliptical&7793&8.3&22.0&17.1&18.5&0.27&0.95&45\\

2141$+$279&Elliptical&8530&21.3&23.4&16.7&25.6&0.0003&0.74&148\\

\tableline
{\bf RLQ}&&&&&&&&&\\
0137$+$012&Elliptical&5093&13.1&22.6&17.2&17.3&0.8&0.85&35\\

0736$+$017&Elliptical&8909&13.3&22.9&16.9&16.2&1.9&0.97&13\\

1004$+$130&Elliptical&501&8.2&21.5&16.9&15.0&5.8&0.94&29\\

2141$+$175&Elliptical&1381&3.7&20.3&17.2&16.0&3.2&0.55&118\\

2247$+$140&Elliptical&8092&10.7&22.4&17.2&16.9&1.3&0.63&118\\

2349$-$014&Elliptical&13463&18.1&22.7&15.9&16.0&0.9&0.89&45\\
\tableline

\tableline
{\bf RQQ}&&&&&&&&&\\
0054$+$144&Elliptical&6050&8.1&21.7&16.6&15.5&2.7&0.61&108\\

0157$+$001&Elliptical&12826&7.9&20.9&15.8&16.2&0.72&0.83&97\\

0244$+$194&Elliptical&2744&18.9&22.7&17.5&16.8&1.9&0.92&77\\

0257$+$024&Disc&15361&10.1&21.5&15.9&19.5&0.04&0.89&134\\

0923$+$201&Elliptical&1733&8.2&22.1&17.2&15.7&4.2&0.98&141\\

0953$+$415&Elliptical&91&7.1&22.4&18.2&15.2&15.4&0.86&115\\

1012$+$008&Elliptical&1056&23.0&23.8&16.6&16.2&1.5&0.64&109\\

1635$+$119&Elliptical&34765&6.3&21.6&16.8&18.1&0.3&0.69&179\\

2344$+$184&Disc&22348&8.8&22.7&17.2&19.2&0.2&0.79&146\\
\tableline
\end{tabular}
\caption{\small The outcome of attempting to model the AGN host galaxies as
either an exponential disc or a de Vaucouleurs spheroid.
The preferred host galaxy morphology is given in column 2, 
with the $\Delta\chi^{2}$ between the chosen and alternative model 
given in column 3.  In column 4\, $r_{1/2}$ is given irrespective of the 
chosen host morphology.
Column 5 lists $\mu_{1/2}$ in units of $R$ mag arcsec$^{-2}$.
Columns 6 and 7 list the integrated apparent magnitudes of the 
host galaxy and fitted nuclear component converted from F675W to 
Cousins $R$-band, 
while column 8 gives 
the ratio of integrated galaxy and nuclear luminosities.  Columns 
9 and 10 give the axial ratio and position angle of the best-fit host 
respectively.}
\end{table}

\clearpage
\begin{table}
\begin{tabular}{clcrccc}
\tableline
\tableline
Source & Host& $\beta$&$\Delta\chi^{2}$ &$R_{host}$ & $R_{nuc}$  & $L_{nuc}/L_{host}$  \\

\tableline
{\bf RG}&&&&&&\\
0345$+$337&Elliptical&0.249&1.75&18.0&21.1&0.06\\

0917$+$459&Elliptical&0.229&263.63 &16.0&19.4&0.04\\

0958$+$291&Elliptical&0.253&16.03 &17.1&18.5&0.27\\

2141$+$279&Elliptical&0.246&2.16 &16.7&26.1&0.0002\\

\tableline
{\bf RLQ}&&&&&&\\
0137$+$012&Elliptical& 0.185&126.34 &17.0&17.4&0.72\\

0736$+$017&Elliptical& 0.193&238.61 &16.8&16.2&1.64\\

1004$+$130&Elliptical& 0.253&4.65&16.9&15.0&5.83\\

2141$+$175&Elliptical& 0.280&22.95&17.3&16.0&3.44\\

2247$+$140&Elliptical& 0.249&16.62&17.2&16.9&1.31\\

2349$-$014&Elliptical&0.258&9.55& 16.0&16.0&0.93\\

\tableline
{\bf RQQ}&&&&&&\\
0054+144&Elliptical&0.251&2.89 &16.6&15.5&2.75\\

0157+001&Elliptical&0.238&133.18 &15.8&16.2&0.68\\

0244+194&Elliptical&0.220&47.42 &17.5&16.8&1.80\\

0257+024&Disc&0.754&2850.72 &15.9&19.5&0.04\\

0923+201&Elliptical&0.299&43.99 &17.3&15.7&4.60\\

0953+415&Elliptical&0.266&8.77&18.2&15.2&15.8\\

1012+008&Elliptical&0.377&101.60 &16.8&16.2&1.81\\

1635+119&Elliptical&0.183&549.58 &16.7&18.3&0.23\\

2344+184&Disc&0.428&1043.76 &17.0&19.2&0.12\\
\tableline
\end{tabular}
\caption{\small The outcome of the variable-$\beta$ modelling.
Column 2 lists the host morphology of the best fitting `fixed $\beta$' model 
(results of which are given in table 2). 
The best-fitting values for the $\beta$ profile parameter are given in 
column 3.  The $\Delta\chi^{2}$ of column 4 quantifies the improvement
in fit between this variable-$\beta$ model and the best-fitting disc or 
elliptical model. As with table 2, columns 5 and 6 
give the integrated apparent magnitudes of the host galaxy and nuclear 
component converted to the Cousins $R$ filter.  
Column 7 lists the ratios of nuclear and integrated host luminosity.}
\end{table}

\clearpage

\begin{table}
\begin{tabular}{ccccc}
\tableline
\tableline
Source & $M_{R}(host)$ &$M_{R}(nuc)$&$(R-K)_{host}$&$(R-K)_{nuc}$\\

\tableline
{\bf RG}&&&&\\
0345$+$337&$-$23.05&$-$19.63&3.23&5.32\\

0917$+$459&$-$24.20&$-$20.65&2.83&4.49\\

0958$+$291&$-$23.25&$-$21.70&3.03&3.97\\

2141$+$279&$-$24.09&$-$14.94&3.02&9.85\\

\tableline
{\bf RLQ}&&&&\\
0137$+$012&$-$24.04&$-$23.53&2.81&3.15\\

0736$+$017&$-$23.58&$-$24.01&3.15&2.73\\

1004$+$130&$-$24.10&$-$25.70&2.96&2.22\\

2141$+$175&$-$23.51&$-$24.52&2.53&1.97\\

2247$+$140&$-$23.80&$-$23.80&2.81&2.60\\

2349$-$014&$-$24.32&$-$24.00&2.86&3.64\\

\tableline
{\bf RQQ}&&&&\\
0054$+$144&$-$23.62&$-$24.49&3.14&1.32\\

0157$+$001&$-$24.29&$-$23.70&2.94&2.82\\

0244$+$194&$-$22.77&$-$23.24&2.54&2.72\\

0257$+$024&$-$23.32&$-$19.69&2.70&4.66\\

0923$+$201&$-$23.25&$-$24.57&3.36&2.66\\

0953$+$415&$-$22.82&$-$25.52&2.84&2.59\\

1012$+$008&$-$23.78&$-$23.95&3.78&1.72\\

1635$+$119&$-$23.05&$-$21.54&3.27&3.23\\

2344$+$184&$-$22.54&$-$20.30&3.11&4.66\\

\tableline

\end{tabular}
\caption{\small Absolute magnitudes $M_R$, and $R-K$ colours of the best-fitting 
host galaxy and nuclear component for each AGN.
Columns 2 and 3 give the $R$-band absolute magnitudes ($M_R$) 
derived from the current modelling of the HST data, 
assuming a spectral index of 
$\alpha=1.5$ (where $f_{\nu}\,\propto\,\nu^{-\alpha}$) for the galaxy
and $\alpha=0.2$ for the quasar.
Columns 4 and 5 list the $R-K$ colours of the host galaxy and nuclear
component respectively. These colours were derived by combining 
12-arcsec aperture $R$-band photometry from our HST-based models 
with the 12-arcsec 
aperture $K$-band photometry derived by Taylor {\it et al.} (1996), to
minimize the uncertainty introduced by errors in constraining the galaxy
scalelengths at $K$.}
\end{table}

\begin{table}
\begin{tabular}{crrrrc}
\tableline
\tableline
Source & ${\rm m_{sph}/10^{11} M_{\odot}}$ & ${\rm m_{bh}^a/10^9
M_{\odot}}$ & ${\rm m_{bh}^b/10^9
M_{\odot}}$ & ${\rm m_{bh}^c/10^9 
M_{\odot}}$&$M_{R}$ (Eddington) \\

\tableline
{\bf RG}&&&\\
0345$+$337 & 6.3\phantom{0}  & 5.0\phantom{0}  & 20\phantom{.00} &    & $-$24.8\\
0917$+$459 & 22\phantom{.00} & 33\phantom{.00} & 25\phantom{.00} & 25\phantom{.00} & $-$28.3\\    
0958$+$291 & 7.8\phantom{0}  &  6.9\phantom{0} & 17\phantom{.00} & 34\phantom{.00} & $-$25.4\\  
2141$+$279 & 20\phantom{.00} & 27\phantom{.00} & 16\phantom{.00} & 19\phantom{.00} & $-$27.9\\    

\tableline
{\bf RLQ}&&&\\
0137$+$012  & 18\phantom{.00} &  25\phantom{.00} & 17\phantom{.00} & 58\phantom{.00} &$-$27.8\\  
0736$+$017  & 11\phantom{.00} &  12\phantom{.00} & 22\phantom{.00} &    &$-$26.4\\  
1004$+$130  & 20\phantom{.00} &  28\phantom{.00} & 13\phantom{.00} & 15\phantom{.00} &$-$28.0\\  
2141$+$175  & 10\phantom{.00} &  11\phantom{.00} & 12\phantom{.00} &    &$-$26.2\\  
2247$+$140  & 14\phantom{.00} &  17\phantom{.00} & 18\phantom{.00} &    &$-$27.1\\  
2349$-$014  & 25\phantom{.00} &  40\phantom{.00} & 12\phantom{.00} & 45\phantom{.00} &$-$28.6\\  

\tableline
{\bf RQQ}&&&\\
0054$+$144 &  12\phantom{.00}  & 13\phantom{.00}& 0.9\phantom{0}   & 3.7\phantom{0} & $-$26.5\\ 
0157$+$001 &  24\phantom{.00}  & 38\phantom{.00}& 2.2\phantom{0}   & 11\phantom{.00}  & $-$28.5\\    
0244$+$194 &  4.6\phantom{0}   & 3.2\phantom{0} & $<$0.6\phantom{0}& $<$2.3\phantom{0} & $-$24.0\\    
0257$+$024 &  3.2\phantom{0}   & 1.8\phantom{0} & 1.2\phantom{0}   & 4.2\phantom{0} & $-$22.9\\    
0923$+$201 &  7.8\phantom{0}   & 6.9\phantom{0} & $<$0.8\phantom{0}& $<$3.0\phantom{0} & $-$25.4\\    
0953$+$415 &  4.9\phantom{0}   & 3.4\phantom{0} & $<$0.8\phantom{0}& $<$3.1\phantom{0} & $-$24.1\\    
1012$+$008 &  14\phantom{.00}  & 16\phantom{.00}& 1.0\phantom{0}   & 4.2\phantom{0} & $-$26.2\\    
1635$+$119 &  6.3\phantom{0}   & 5.0\phantom{0} & 2.5\phantom{0}   & 12\phantom{.00}  & $-$24.8\\    
2344$+$184 &  2.9\phantom{0}   & 1.6\phantom{0} & $<$0.5\phantom{0}& $<$1.8\phantom{0} & $-$22.6\\

\tableline

\end{tabular}
\caption{\small The results of calculating spheroid mass, and hence black
hole mass/luminosity from host-galaxy spheroid luminosity.
Column 2 gives the derived spheroid mass 
(in units of $10^{11} {\rm M_{\odot}}$), calculated from the
values of $M_R$ given in Table 4
using the mass:light ratio relation given by Magorrian {\it et al.} (1998).
Column 3 then gives the estimated mass of the central black hole
(in units of $10^{9} {\rm M_{\odot}}$) derived from the data in column 2
using the high mass form of the 
$m_{sph}:m_{bh}$ relation deduced by Magorrian {\it et al.} 
(1998), while columns 4 and 5 give alternative estimates of $m_{bh}$ based
on $P_{5GHz}^{total}$ and $P_{5GHz}^{core}$
Finally column 6 gives the predicted Eddington 
$M_R$ corresponding to $m_{bh}^a$ given in column
3.}
\end{table}

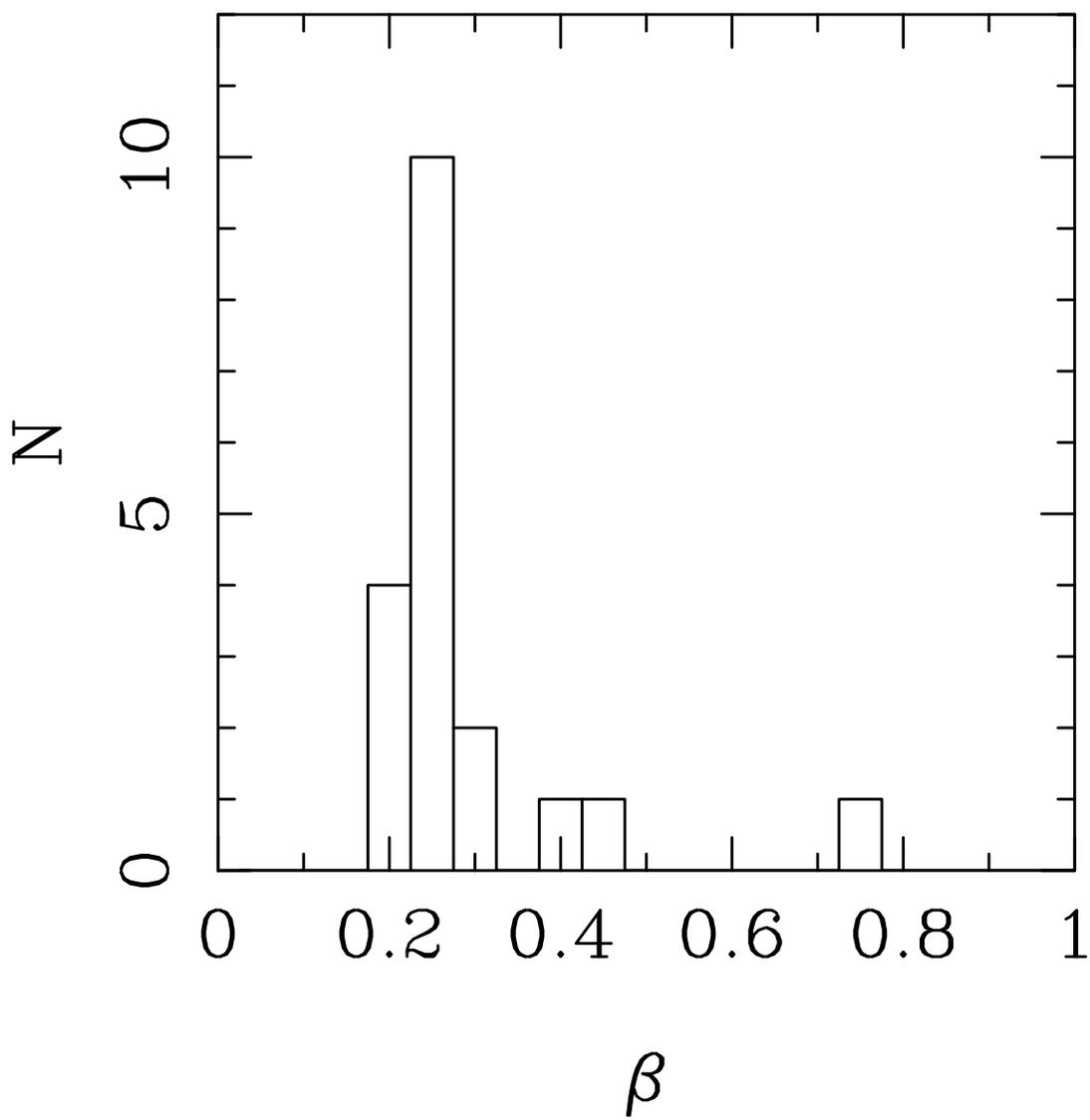
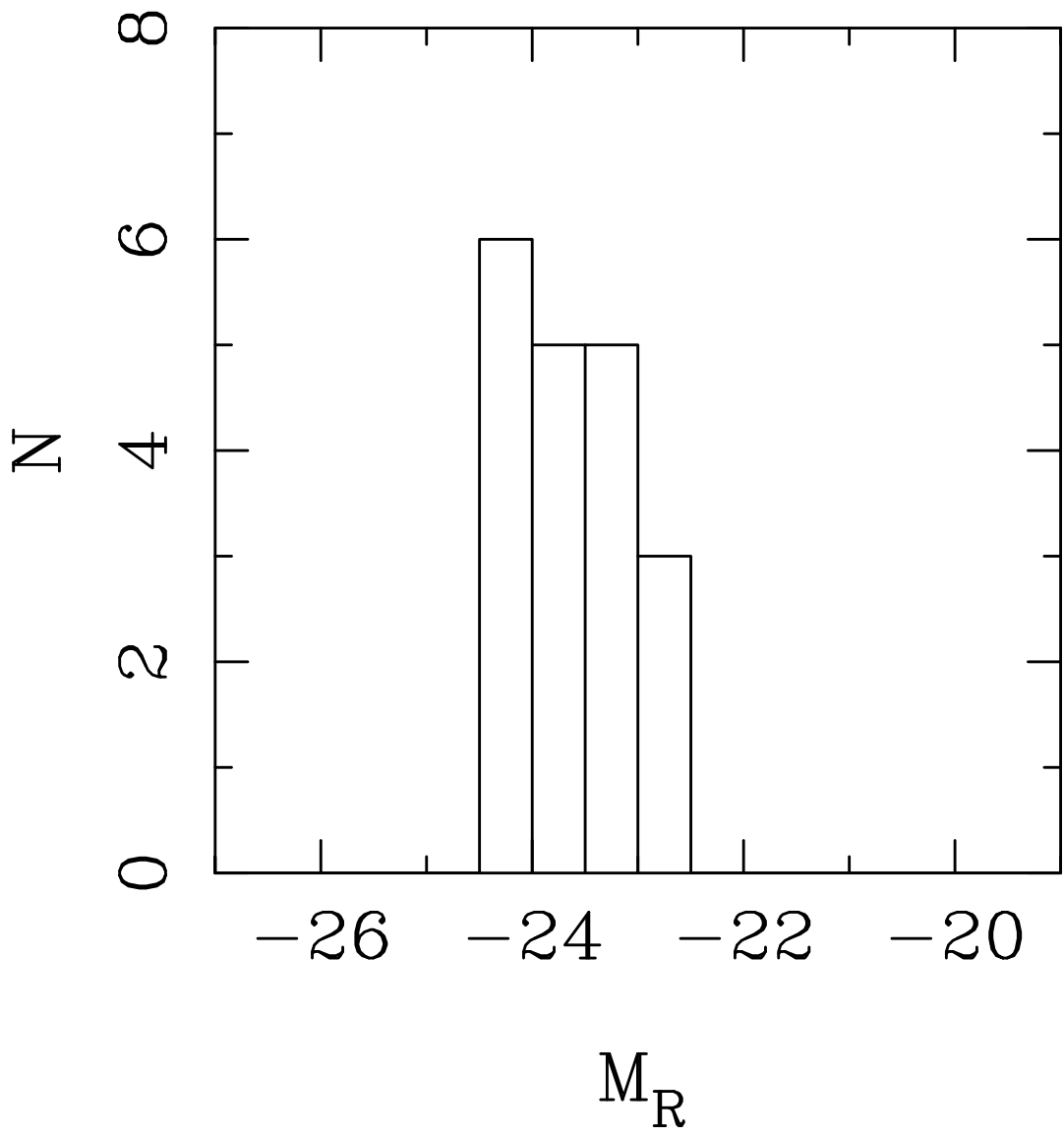
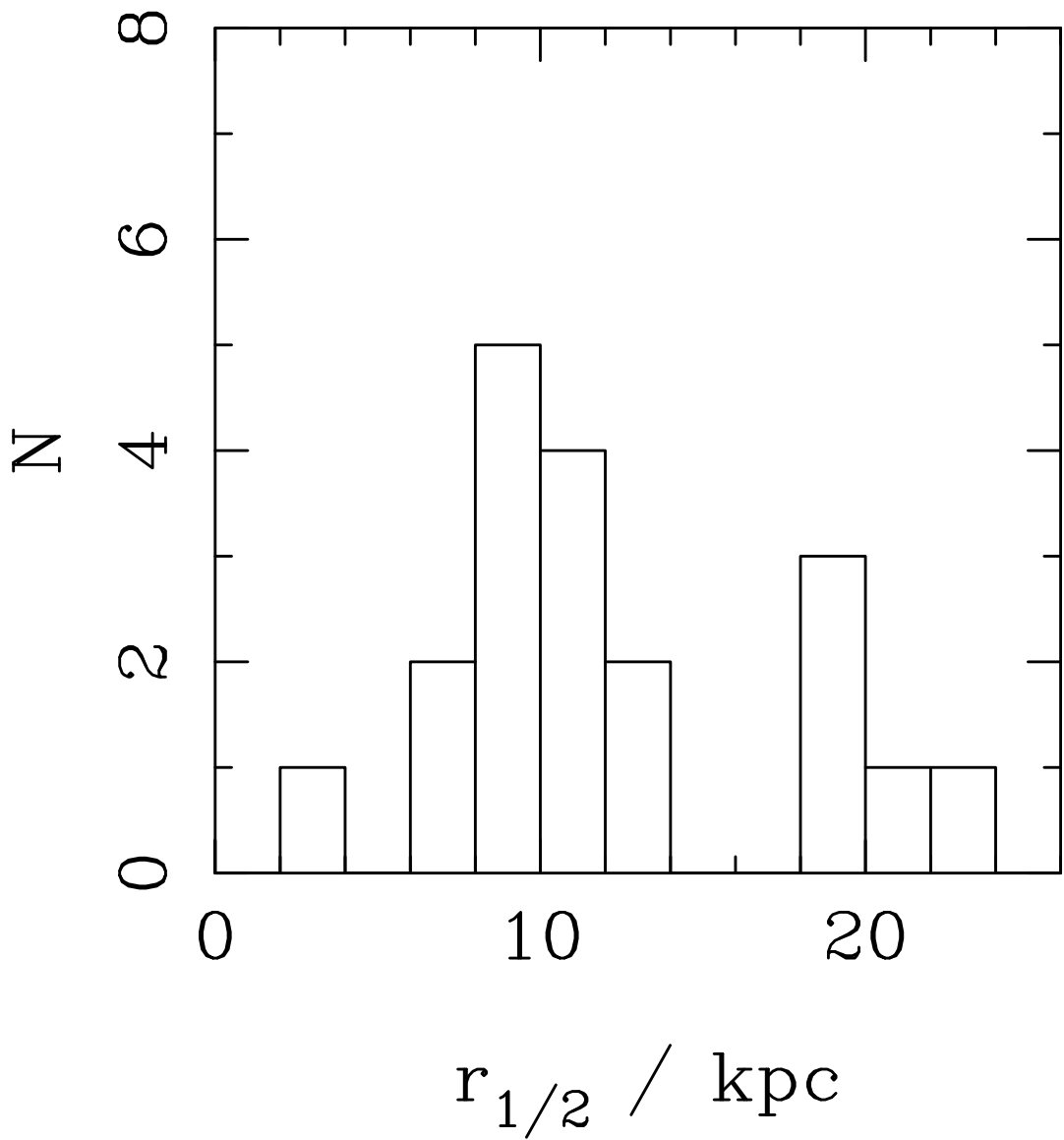
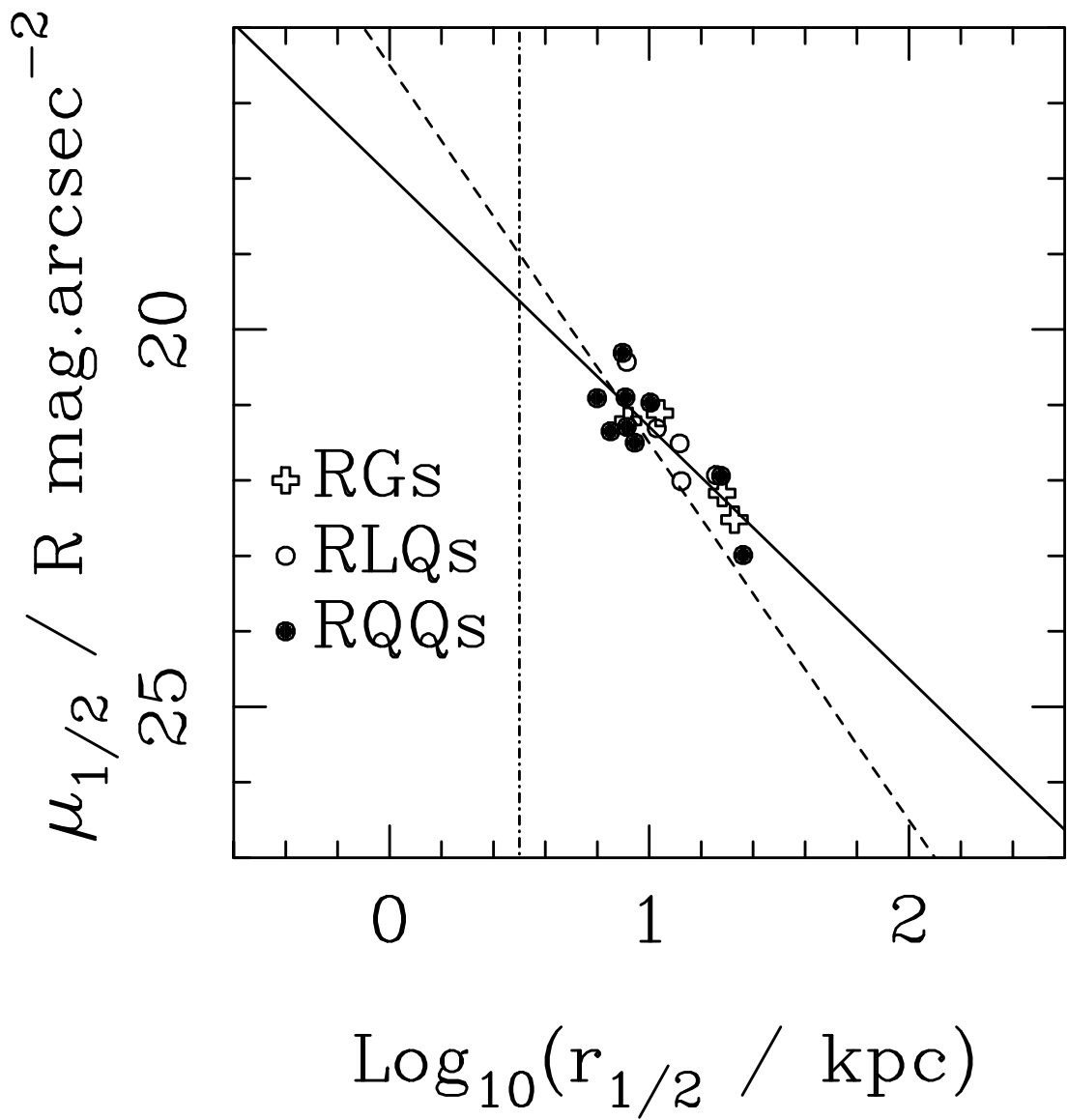
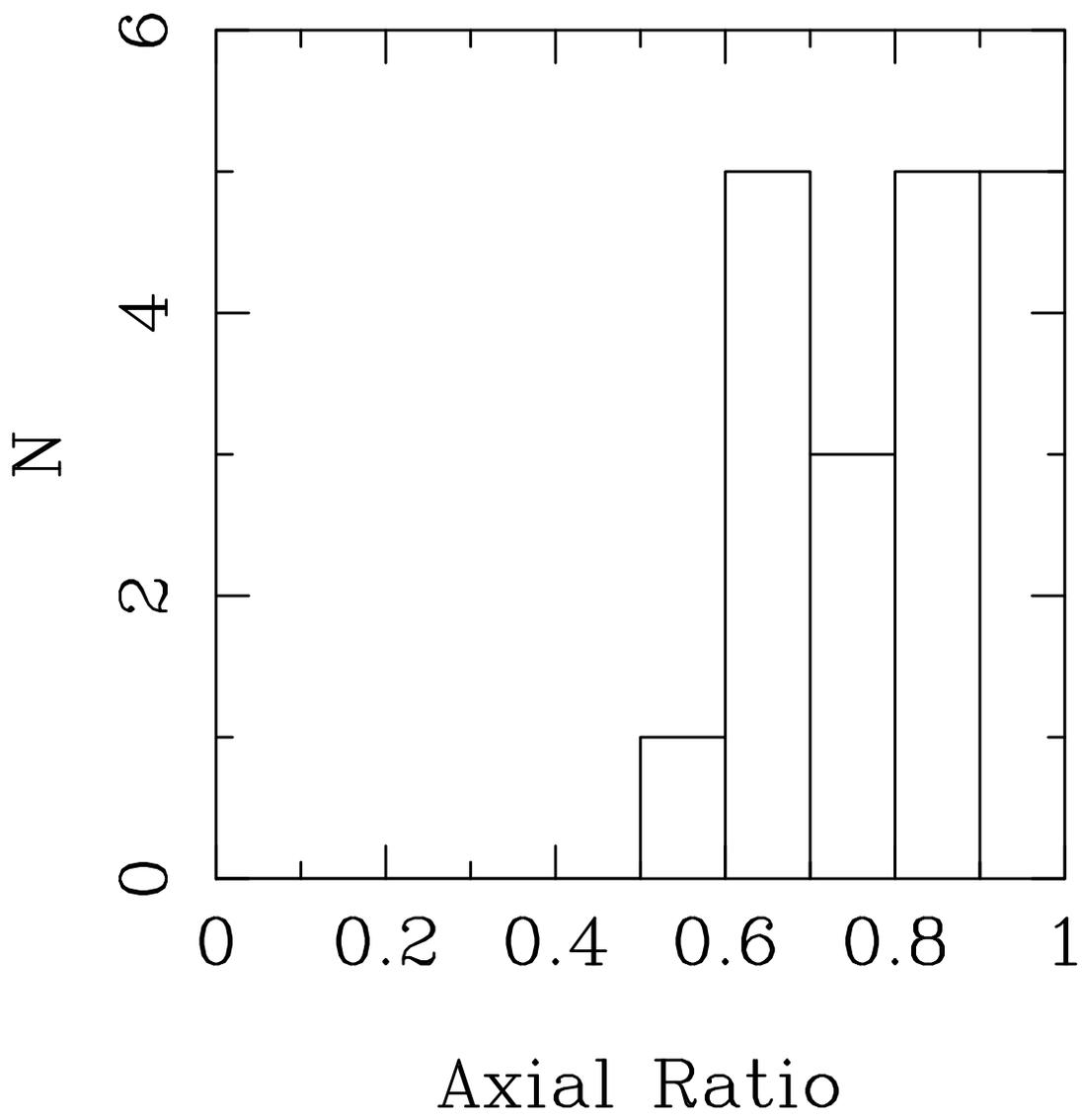
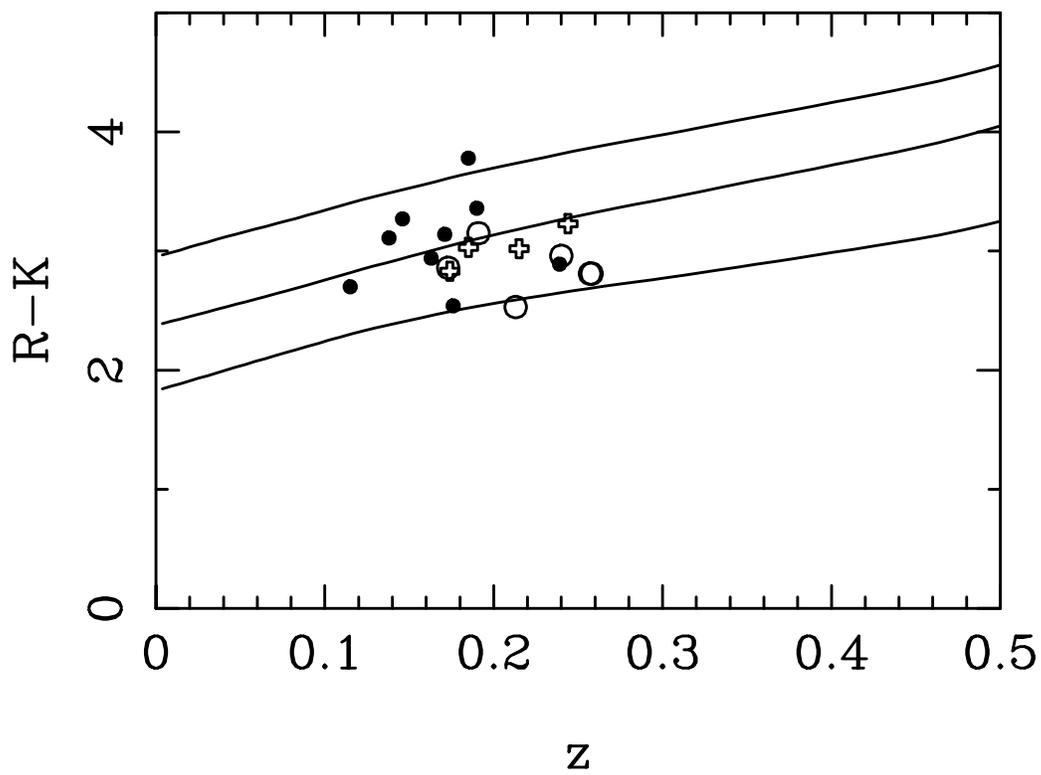
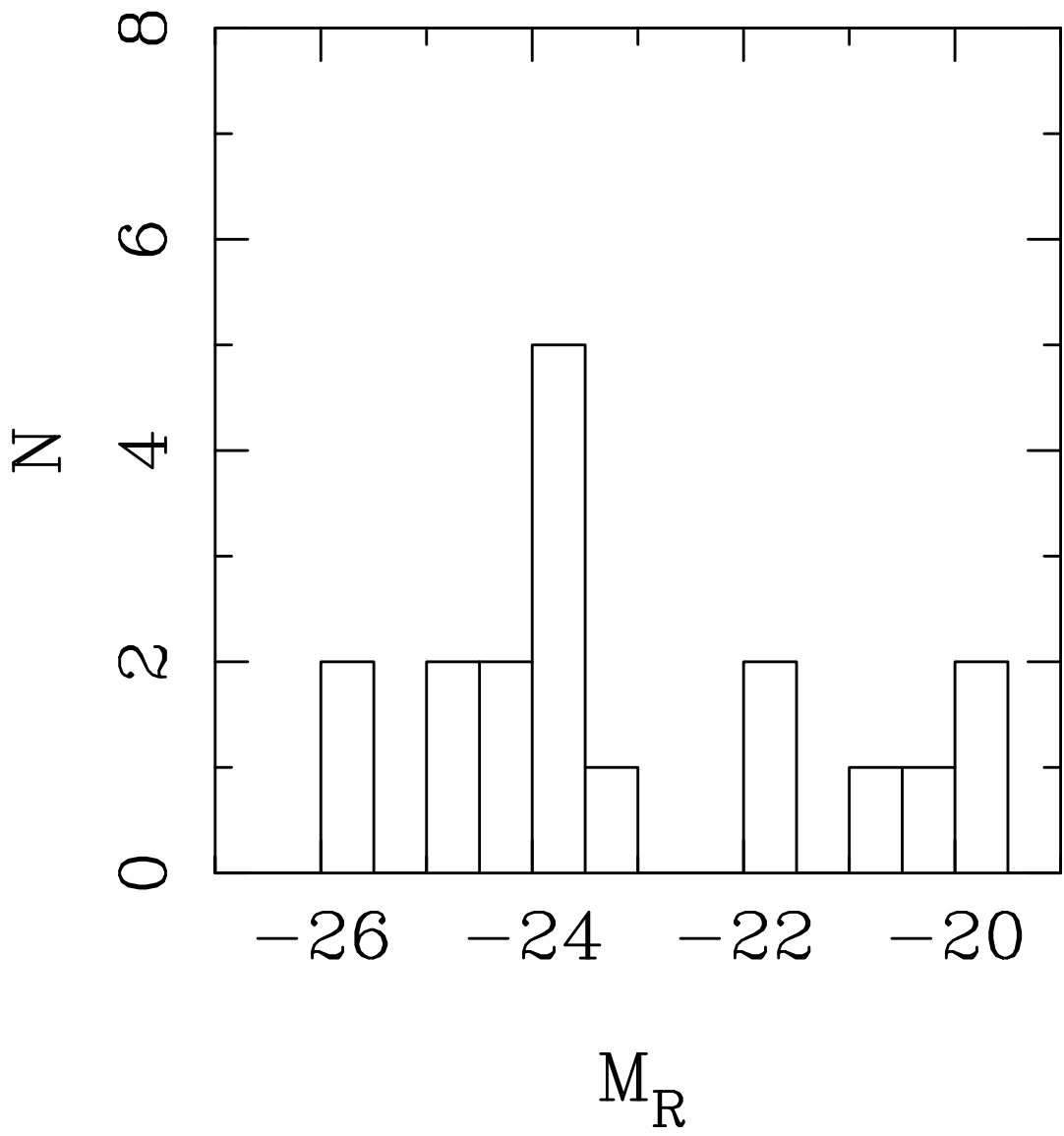
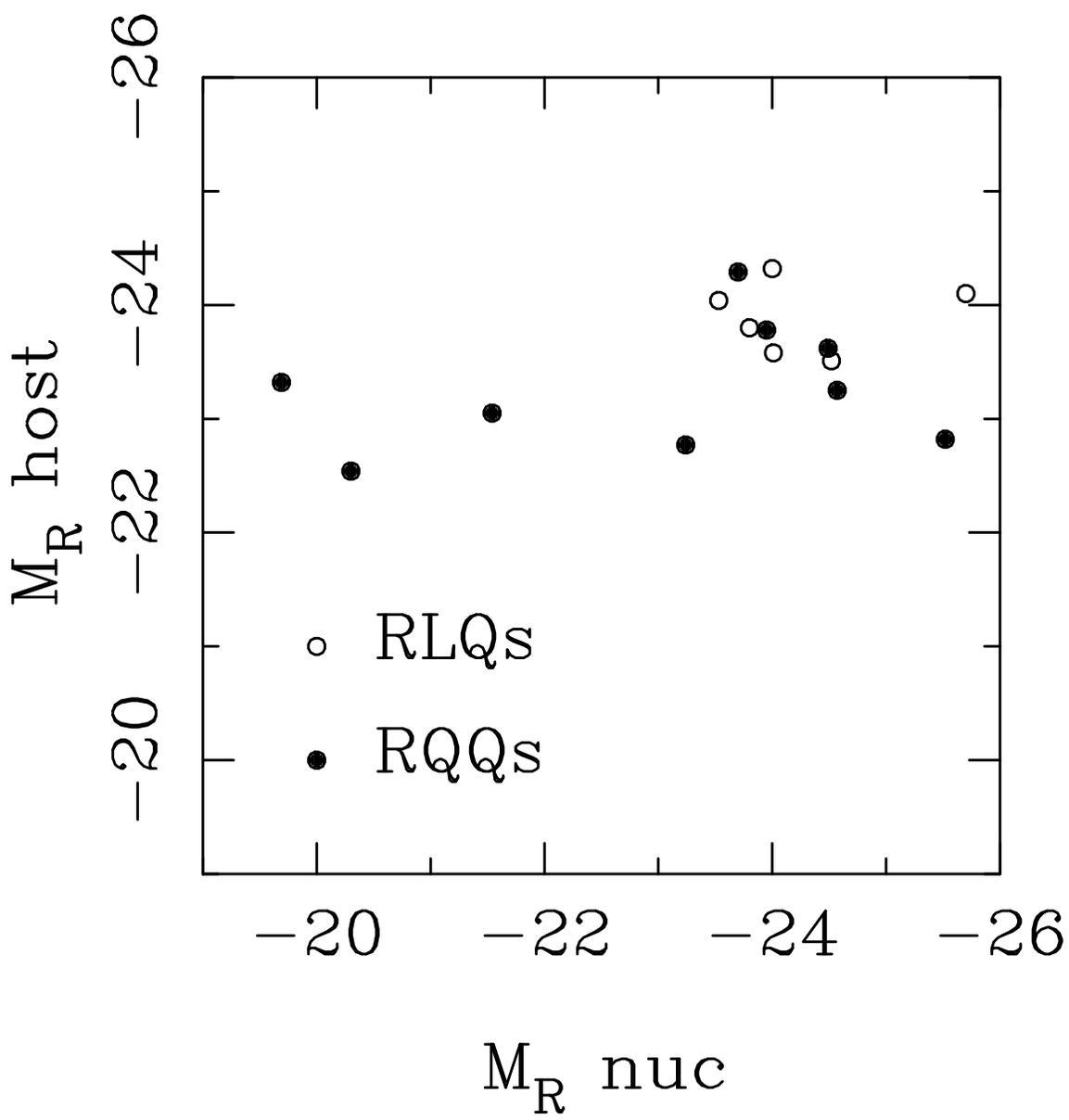
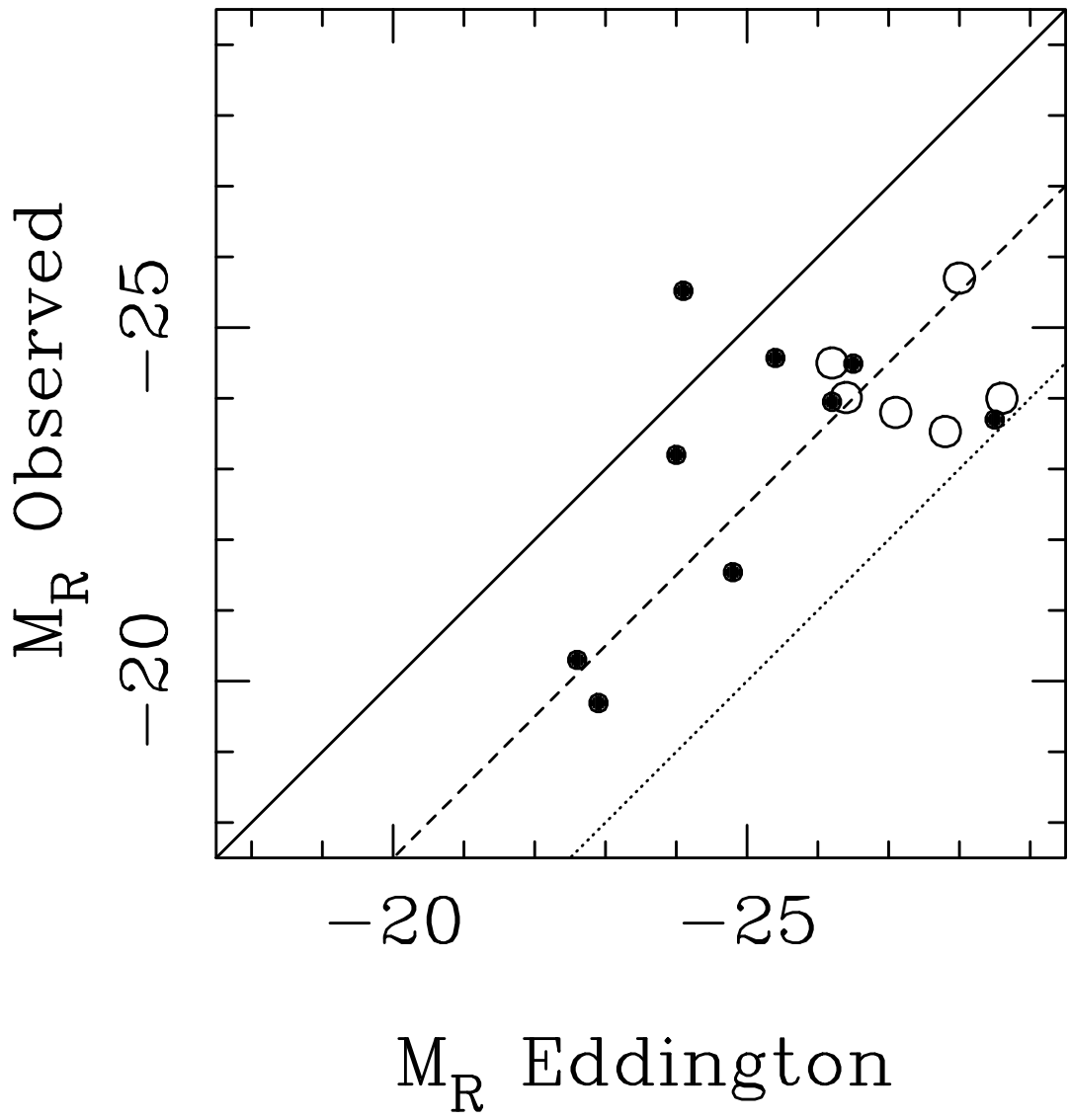
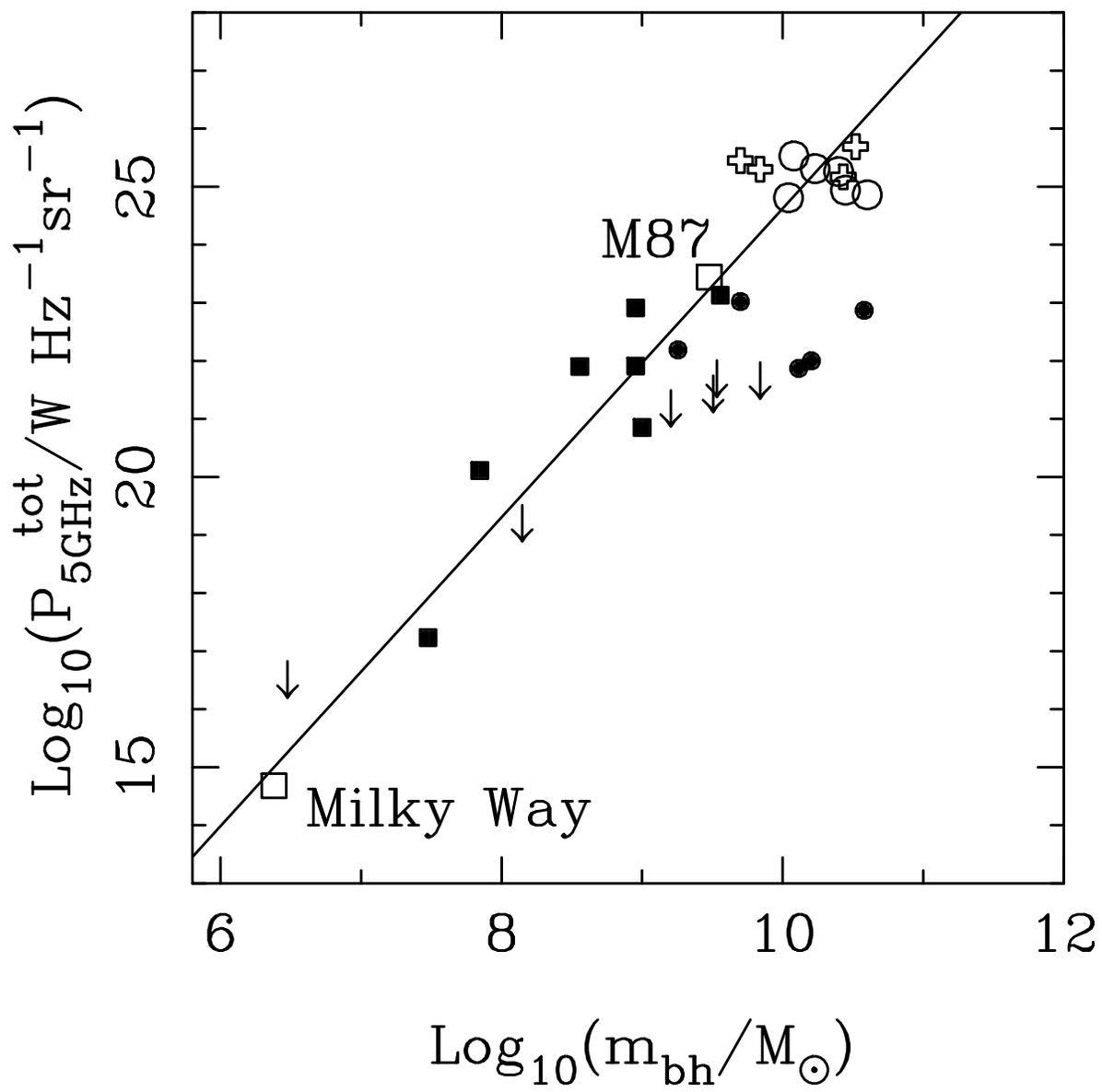
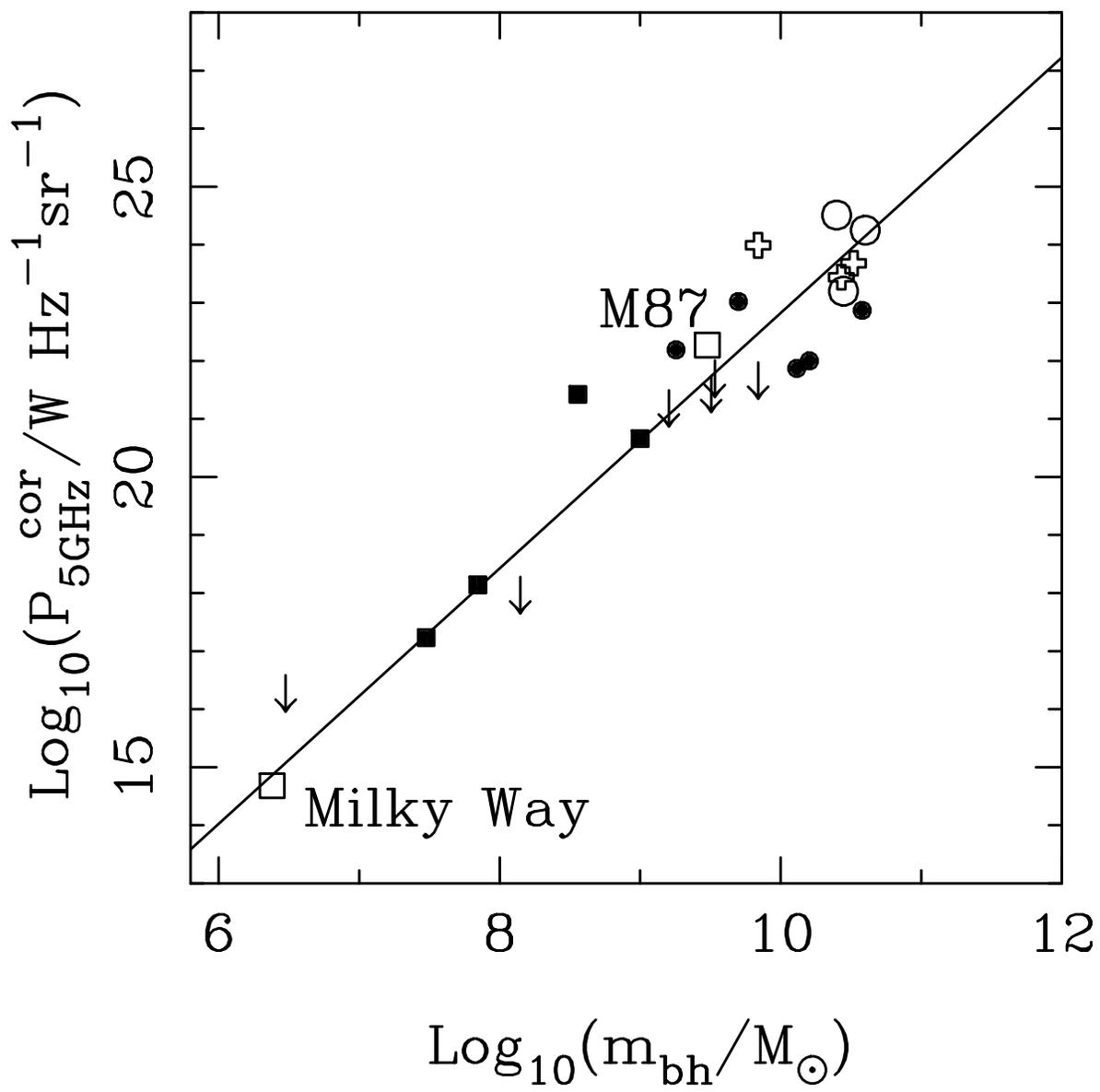
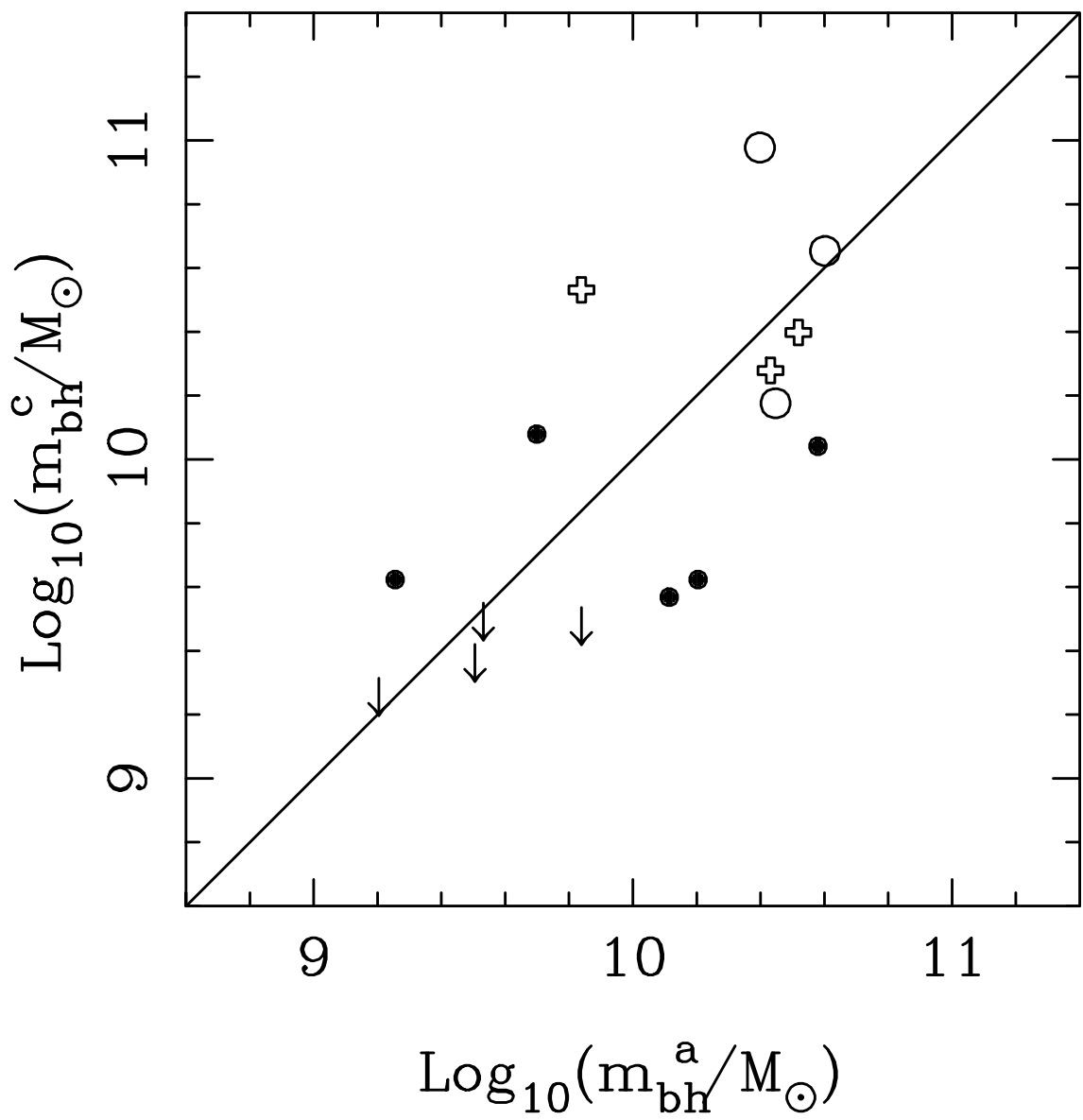
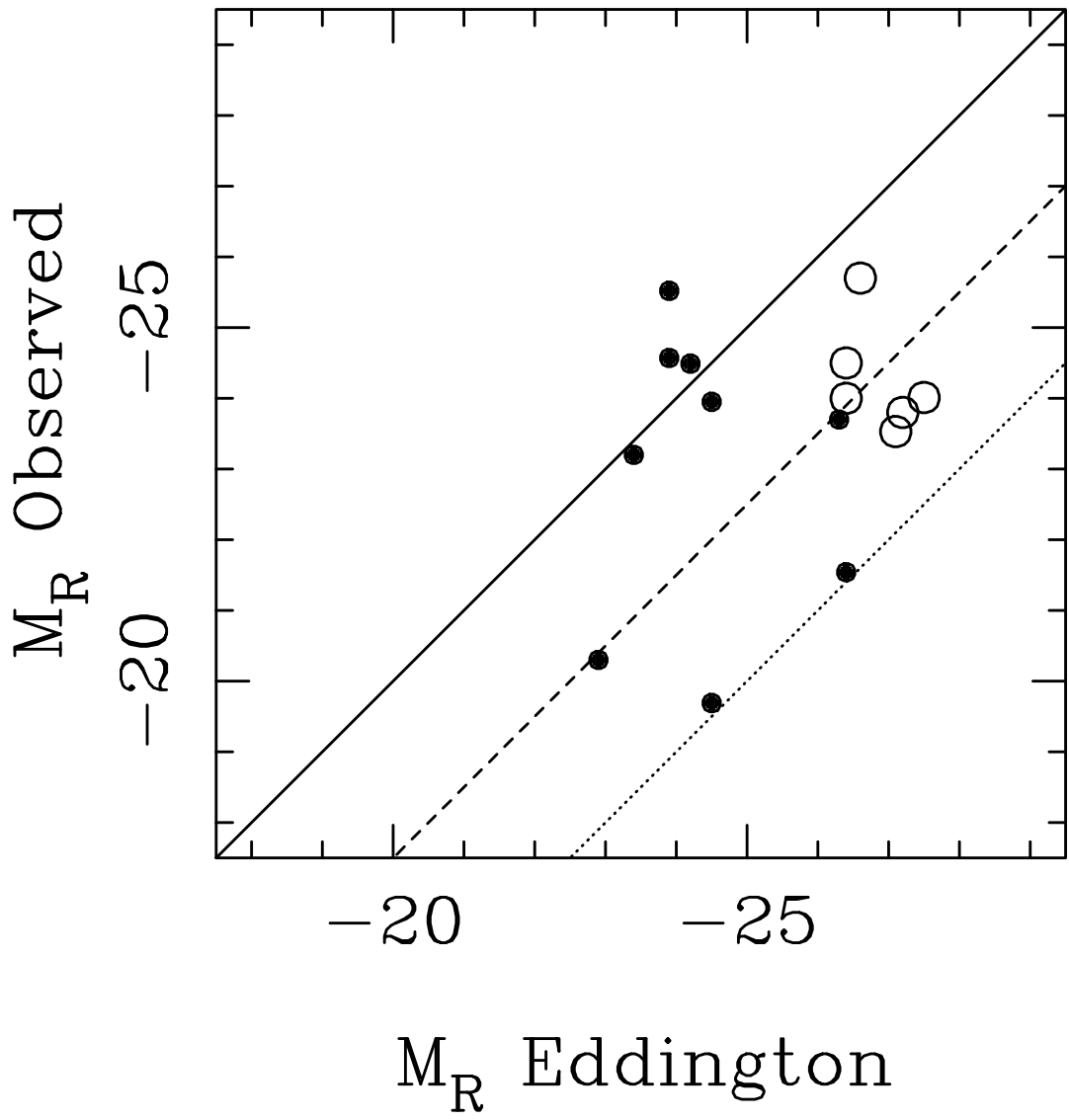
\end{document}